\documentstyle{article}
\begin{document}

\begin{center}

{\LARGE \bf Miscellaneous photometric variations in cataclysmic variables: 
\vspace{1ex}

V455~And, SS~Cyg, AQ~Men, LQ~Peg, RW~Tri 
\vspace{1ex}

and UX~UMa}\footnote{Based 
partially on observations taken at the Observat\'orio do Pico dos Dias / LNA}

\vspace{1cm}

{\Large \bf Albert Bruch}

\vspace{0.5cm}
Laborat\'orio Nacional de Astrof\'{i}sica, Rua Estados Unidos, 154, \\
CEP 37504-364, Itajub\'a - MG, Brazil
\vspace{1cm}

(Published in: New Astronomy, Vol.\ 78, 101369 (2020))
\vspace{1cm}
\end{center}

\begin{abstract}
Cataclysmic variables are among the photometrically most unstable stars in
the zoo of stellar objects, exhibiting light variations on all time-scales
between millennia and seconds. The literature is full of reports on variable
phenomena which often require independent confirmation before they can be 
accepted as established facts. In this contribution I investigate accounts 
on miscellaneous variable features observed in six cataclysmic variables, 
drawing for this purpose largely on archival data, most of which have not been
investigated in detail in the past, and complementing these data with some
new observations. This enabled to confirm and expand upon some hitherto 
unconfirmed features in the light curves of these star, as well as the 
rejection of some others, while in still other cases an unambiguous answer 
to questions arising from previous papers was not possible.

\phantom{.}

{\parindent0em Keywords:
Stars: Binaries: Close -- 
Stars: Novae, Cataclysmic variables -- 
Stars: Dwarf novae --
Stars: Individual: V455~And; SS~Cyg; AQ~Men; LQ~Peg; RW~Tri; UX~UMa}
\end{abstract}

\section{Introduction}
\label{Introduction}

Among the many types of variable stars the cataclysmic variables (CVs)
arguably are the champions of variability. CVs are interacting binary systems
consisting of a Roche-lobe filling late type star (the secondary) orbiting
a white dwarf (the primary) and transferring matter to the latter which forms
an accretion disk before it settles onto the compact object (unless the
magnetic field of the primary is strong enough to guide it directly to its
surface). For a thorough review of the properties of CVs, see, e.g.,
Warner (1995). Variable phenomena occur in these systems periodically,
quasi-periodically or irregularly on all times-scales
from millennia to decades (outbursts of classical and recurrent novae),
years, month and weeks (high and low states, dwarf nova outbursts),
days and hours (orbital variations, superhumps and beat phenomena between
them), minutes and seconds (white dwarf rotation, flickering, quasi-periodic
oscillations, dwarf nova oscillations).

Their thorough observation and characterization permits to elucidate many
aspects of the nature of CVs in general as well as in individual systems.
The literature is, however, full of reports on details which remain
unconfirmed and may or may not represent true features in the investigated
stars. If they are not, taking them at face values easily leads to confusion
and miss-interpretations. It is therefore worthwhile to use additional data
in order to verify unconfirmed claims of variable features observed at one 
instance or the other. Fortunately, a lot of such and so far little or not
studied data are available in public data arquives.

Using such data and complementing them with some new dedicated observations,
I investigate in this small contribution miscellaneous variations in six
CVs (some better and some less well known) in order to verify previous
unconfirmed features and to add some new details concerning
their temporal behaviour on time-scales of seconds to weeks.
V455~And is an intermediate polar. Here, the orbital period, the
occurrence of superhumps, and the temporal behaviour of high
frequency oscillations are investigated (Sect.~\ref{V455 And}). SS~Cyg
(Sect.~\ref{SS Cyg}) is the prototype of the dwarf nova class of CVs. Some
details about its orbital modulations are studied along with variations at
higher frequencies. Orbital variations and
a possible modulation with a period of some days in the novalike variable
AQ~Men are looked at in Sect.~\ref{AQ Men}, while controversies in the
literature about variations on hourly time-scales in the VY~Scl star
LQ~Peg are discussed in Sect.~\ref{LQ Peg}. A follow-up on recent reports
of superhumps and of oscillations on time-scales of weeks in RW~Tri is
presented in Sect.~\ref{RW Tri}. Finally, in Sect.~\ref{UX UMa} it is
verified if superhumps observed in UX~UMa in 2015 are also present at
other epochs.

\section{The data}
\label{The data}

The data used in this study consist of light curves which in most cases
span several hours and have a time resolution of (with a few exceptions)
better than one minute
(down to 5~s). Most of them were retrieved either from the AAVSO
(American Association of Variable Star Observers) International Database
or from the Centre de Donn\'ees Astronomique de Strassbourg (CDS). Only for
AQ~Men new and so far unpublished observations are used. Details about the
data are given in the sections concerning the individual objects. Whenever
light curves of different nights were combined, the time stamps of all data
points were first transformed into barycentric Julian Date on the
Barycentric Dynamical Time scale, using the online tool of Eastman et al.\
(2010) in order to remove any light travel time differences
in the solar system. This was not necessary if the archival data already
included the heliocentric correction (thus neglecting the small difference
between barycentric and heliocentric time). Since the AAVSO data
were observed by a variety of observers with different instruments,
the details of which are not known, the magnitude scale can differ
between the data sets and night-to-night variations seen in the light curves
may not be real. Therefore, unless stated otherwise the average magnitude has 
been subtracted from these light curves before further processing.

\section{V455~Andromedae}
\label{V455 And}

V455~And is a comparatively well studied system. Even so, additional data
still can add to our knowledge of the star, confirming and substantiating
previous observations. Discovered in the Hamburg Quasar Survey
(Hagen et al., 1995) as HS2331+3905, it is considered a DQ~Her type intermediate
polar with a short orbital period of 81.08~min (Araujo-Betancor et al., 2005).
It is also a dwarf nova of the WZ~Sge subtype with only one outburst having
been observed so far (Nogami et al., 2009; Maehara et al., 2009; Matsui et
al., 2009). The orbital light curve is very similar to
that of WZ~Sge. It is double humped and contains a shallow grazing eclipse.
Moreover, Araujo-Betancor et al.\ (2005) detected a positive superhump during
quiescence, while Kozhevnikov (2015) observed a negative superhump. A
puzzling feature is a 3.5~h spectroscopic period detected by
Araujo-Betancor et al.\ (2005) which is in no way related to the orbital period.

In the high frequency regime several periods were first observed in V455~And
by Araujo-Betancor et al.\ (2005) and later studied in more detail by other
authors. A signal on time-scales of 5--6 min which changes slightly from
night to night is interpreted as non-radial pulsations of the white
dwarf. A coherent 1.12 min signal is thought to be the rotation period of
the primary star. A precise period of $P_{\rm spin} = 67.61970396$~s
has been measured by Mukadam et al.\ (2016). Closely related is another,
slightly drifting period of 67.20 s, first mentioned by G\"ansicke (2007)
who interpreted it as being due to the illumination by the rotating white
dwarf of a warped inner accretion disk precessing retrogradely.
Intriguingly, the beat period between the latter and $P_{\rm spin}$ is very
close to the 3.5 h radial velocity variation.

The observational data used for this study were retrieved from
AAVSO International Database. They consist of 116 light curves with a
duration between 77 and 958~min and a time resolution between 20 and 169~s,
all referring to quiescence.
They were obtained in the eleven observing seasons between 2008 and 2018.

\subsection{The orbital period}
\label{V455 And: The orbital period}

\input epsf

\begin{figure}
   \parbox[]{0.1cm}{\epsfxsize=14cm\epsfbox{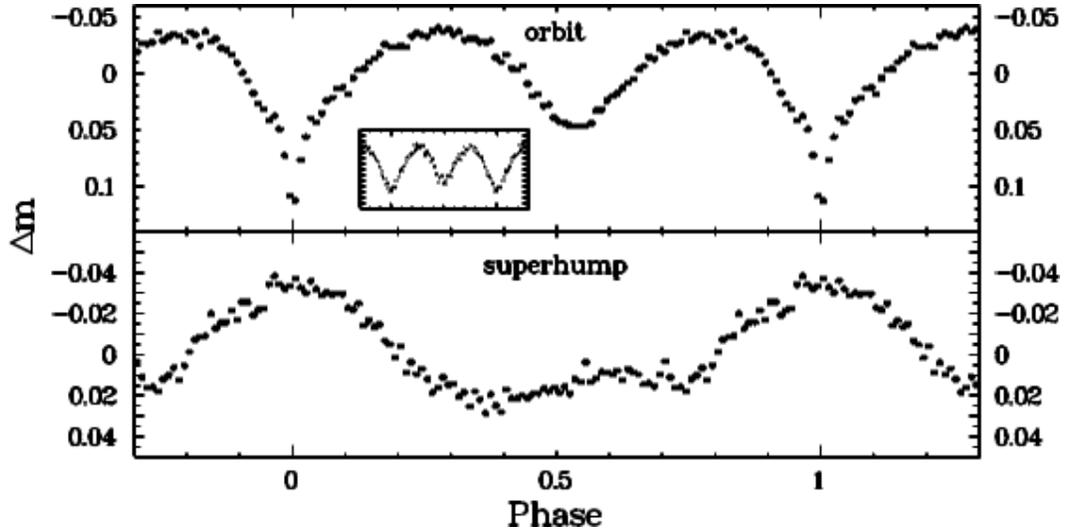}}
      \caption[]{Average waveform of the orbital (top) and superhump
                 (bottom) variations of V455~And. The insert in the upper
                 frame results from folding the light curves on the 1~d$^{-1}$
                 alias of half the orbital period (see text for details).}
\label{v455and-orbit-sh}
\end{figure}

The orbital period of V455~And is known to a rather high accuracy
(Araujo-Betancor et al., 2005). However, the longer time base of the
present data permits an even better determination. For that purpose the light
curves of each observing season between 2008 and 2018 (except 2017 when only
one light curve was observed) were combined and then folded on the orbital
period. This enabled an unambiguous detection of the grazing eclipses which
are not obvious in individual light curves. For each season an eclipse epoch
was determined by fitting a sixth order polynomial to the eclipse in the
folded light curve. The period of Araujo-Betancor et al.\ (2005) is more than
good enough to maintain cycle counts over the entire time base, permitting
thus to associate a cycle number to each of the 10 eclipse epochs. A linear
least squares fit, including also the eclipse epoch informed in eq.~1 of
Araujo-Betancor et al.\ (2005) and weighting each epoch with the total
observing time of each season, then yields the improved ephemeris:
\begin{displaymath}
T_{\rm ecl} = BJD 2454623.52003 (9) + 0.056309188 (2) \times E
\end{displaymath}
where the errors are the formal fit errors. The light curve folded on the
period and binned in phase intervals of width 0.01 is shown in the upper frame
of Fig.~\ref{v455and-orbit-sh}. Not astonishingly, the waveform
is very similar to that shown in Fig.~7 of Araujo-Betancor et al.\ (2005):
the first maximum extends from phase 0 to
0.534 and is thus slighly longer than the second maximum. At 0.084~mag
(measured from its peak to the bottom of the secondary minimum) it has a
slightly lower amplitude than the second maximum which reaches 0.081~mag.
However, it should be noted that the relative amplitudes of the maxima
may change from season to season, as is obvious from fig.~5 of
Kozhevnikov (2015). The eclipse depth, defined here as the magnitude
difference between the secondary minimum and the eclipse bottom, is 0.062~mag.

\subsection{The superhumps}
\label{V455 And: The superhumps}

Claims have been put forward concerning the presence of positive as well
as negative superhumps in V455~And. In observations obtained between 2000
and 2003, Araujo-Betancor et al.\ (2005) observed a periodic modulation
with a period of 83.38~min (or an alias at 88.51~min), slightly longer than the
orbital period, which they interpret as a positive superhump. It is not seen
in observations taken in 2013 and 2014 by Kozhevnikov (2015) who, 
instead, found a modulation with a period of 80.376~min, slightly shorter than
the orbital period. He considers this variation as a negative superhump.

The reality of the positive superhump may be questioned. First, it has a
double humped structure (see Fig.~8 of Araujo-Betancor et al., 2005),
quite unusual for ordinary superhumps in CVs. While there may be some
additional structure in the waveform, the numerous examples shown by
Kato et al.\ (2009) and in other papers of that series clearly show one
dominant maximum during each superhump cycle. More important, however, is
the remarkable proximity of half of the alleged superhump period (41.69~min)
to the 1~d$^{-1}$ alias of the main power spectrum signal (41.72~min, i.e., half 
the orbital period). Folding the present light curves on the 1~d$^{-1}$ alias of
the orbital period, as shown in the insert in the upper frame of
Fig.~\ref{v455and-orbit-sh}, smears out the eclipses, leading to a
waveform with two minima of slightly different depth. This is quite similar
to fig.~8 of Araujo-Betancor et al.\ (2005). Therefore, the corresponding
variations can easily be explained by an aliasing effect and there is no
need to invoke the presence of a permanent superhump during quiescence of
this WZ~Sge star.

\begin{figure}
   \parbox[]{0.1cm}{\epsfxsize=14cm\epsfbox{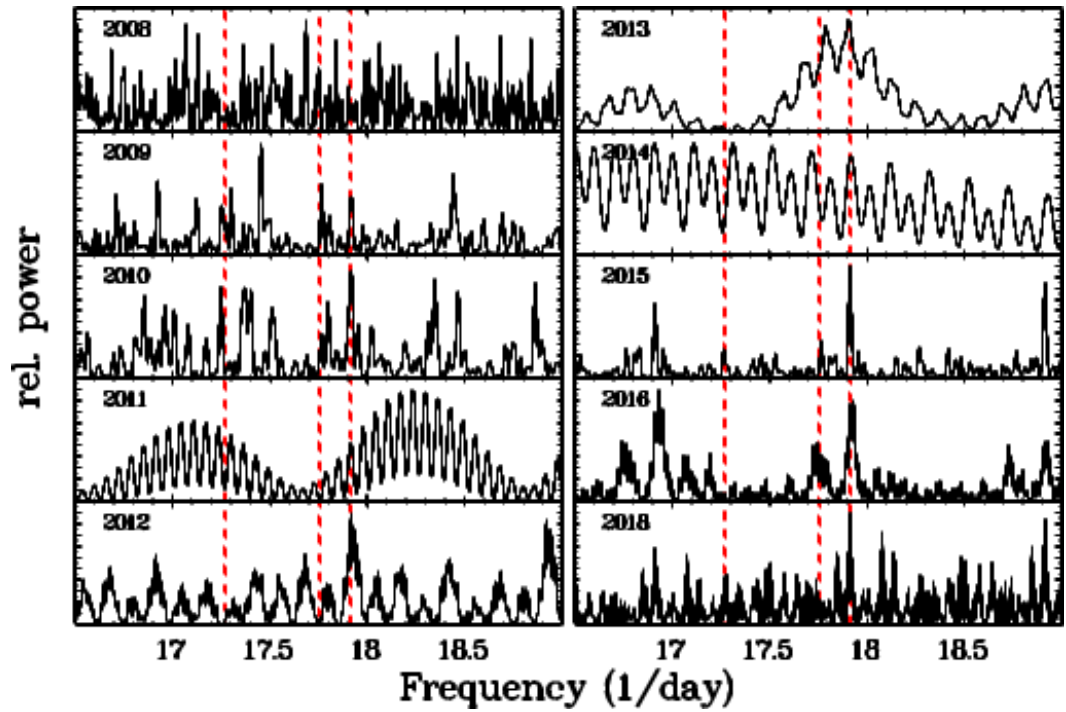}}
      \caption[]{Periodograms of the combined light curves of V455~And
                 during various observing seasons. The broken vertical lines
                 indicate (from left to right) the frequencies of the
                 alleged positive superhump at 83.38~min
                 (Araujo-Betancor et al., 2005), the orbit,
                 and the negative superhump detected by
                 Kozhevnikov (2015).}
\label{v455and-superhump-ps}
\end{figure}

In the present data, a possible positive superhump can at most marginally be
seen during some years,
while the negative superhump is detected during most observing
seasons. This is shown in Fig.~\ref{v455and-superhump-ps} which contains
the Lomb-Scargle periodograms (Lomb, 1976; Scargle, 1982; hereafter also
referred to as power spectra) of the combined
seasonal light curves, noting that those of 2011 and 2014 are based on
only three nights of observations, which explains their peculiar appearance.
The broken vertical lines indicate (from left to right) the frequencies of
the alleged positive superhump (Araujo-Betancor et al., 2005), the
orbit, and the negative superhump (Kozhevnikov, 2015).

While the orbital frequency appears only
weakly during some seasons (remembering that due to the double humped
light curve the bulk of the power is concentrated at twice the orbital
frequency), a significant signal is present at the negative superhump
frequency (and its 1~d$^{-1}$ aliases) in all seasons except 2008 and 2011.
The average of the peak frequencies in the yearly periodograms yields a
superhump period of $80.382 \pm 0.018$~min, entirely consistent with the
value measured by Kozhevnikov (2015).

In order to obtain the waveform of
these variations, a least squares sine fit with the period fixed to half
the orbital period was subtracted from the combined seasonal light curves
in order to remove (at least partly) the orbital variations. The result was
then folded on the superhump period measured during that particular season,
adjusting the phase such that the maximum of the waveform has phase 0. The
average waveform, binned in phase intervals of width 0.01, is shown in the
lower frame of Fig.~\ref{v455and-orbit-sh}. Just as observed by Kozhevnikov
(2015) (see his fig.\ 5) the superhump has two maxima. However, the
intermediate hump at phase 0.6 has a much smaller amplitude in the present
data. The total amplitude of the main hump is $\approx$0.057~mag which is
significantly less than estimated from fig.\ 5 of Kozhevnikov (2015).

Finally, testing if the superhump can
be described by a single period, constant over several years, an attempt was
made to find a value for the period which keeps the superhump maximum in
phase over the entire time base of the data. This turned out not to be
possible, indicating that either the period or the phase (or both) undergo
real variations, however small, over time.

\subsection{The pulsations}
\label{V455 And: The pulsations}

First noticed by Araujo-Betancor et al.\ (2005), multiple signals in the
power spectra of V455~And, corresponding to periods in the 4--6~min range,
have been studied subsequently also by G\"ansicke (2007),
Silvestri et al.\ (2012) and Szkody et al.\ (2013). While most authors 
consider them as due to non-radial pulsations of the white dwarf, 
Szkody et al.\ (2013) question this view based on indications for an origin 
away from the white dwarf photosphere.

None of the cited studies was able to isolate specific frequencies. The
number of significant signals and their frequencies changes from night to
night, only permitting to identify a range for their occurrence, and this
range is not even constant over time. Szkody et al.\ (2013) observed an
evolution of the pulsation period from $\approx$4.2~min in 2009 (two years
after the dwarf nova outburst of V455~And) to $\approx$5.5~min in 2012.
The latter value is similar to that measured by Araujo-Betancor et al.\ (2005)
in 2000 -- 2003 and G\"ansicke (2007) in 2004, i.e., years before
the outburst. Szkody et al.\ (2013)
attribute the change of preferred periods to the cooling of the
white dwarf surface after being heated during the eruption.

\begin{figure}
   \parbox[]{0.1cm}{\epsfxsize=14cm\epsfbox{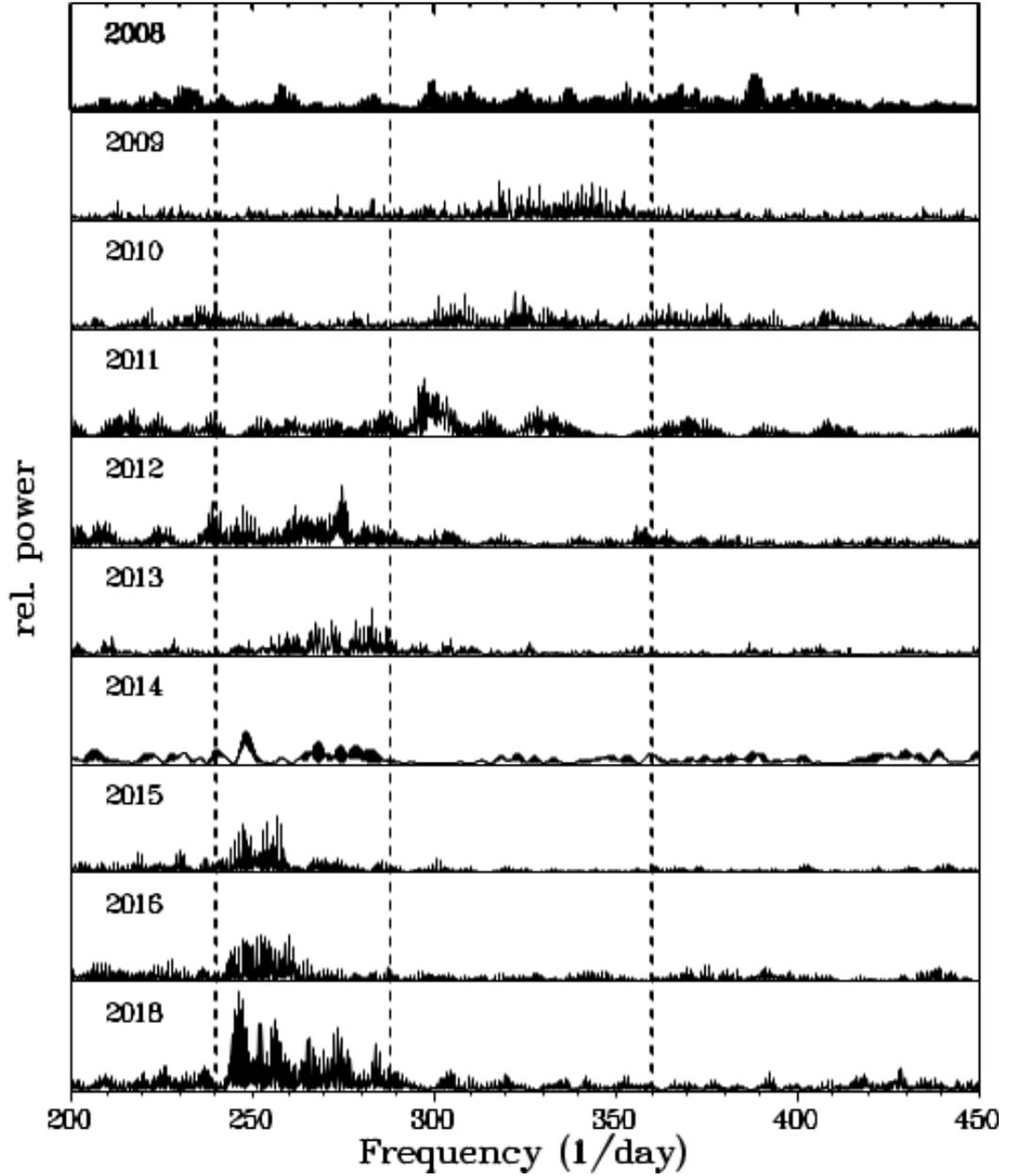}}
      \caption[]{Periodograms of the combined light curves of V455~And
                 during various observing seasons, covering the frequency
                 range of the white dwarf pulsations. All periodograms
                 are drawn on the same vertical scale. The broken vertical
                 lines indicate (from right to left) frequencies corresponding
                 to 4, 5 and 6~min.}
\label{v455and-pulsations}
\end{figure}

The present data permit to document the evolution of the range of
pulsation periods over a longer time base. For this purpose,
Fig.~\ref{v455and-pulsations} shows the relevant frequency interval
of the periodograms calculated from the combined seasonal light curves.
All are drawn on the same scale. If there is a preferred range of
frequencies at all in 2008, it is very broad. However the periodogram
may also be compatible with the absence of well defined pulsations. The
picture becomes clearer in the subsequent years, where a trend for pulsations
from higher to lower frequencies is evident, confirming the findings of
Szkody et al.\ (2013). This trend comes to a halt in 2012, when the
range of frequencies has reached that observed in 2000--2004. Thereafter,
it remains constant until 2018. In line with the explanation for the period
evolution forwarded by Szkody et al.\ (2013), the white dwarf in V455~And
may still have been to hot to permit non-radial pulsations in 2008 and
has reached its equilibrium temperature by 2012.

\begin{figure}
   \parbox[]{0.1cm}{\epsfxsize=14cm\epsfbox{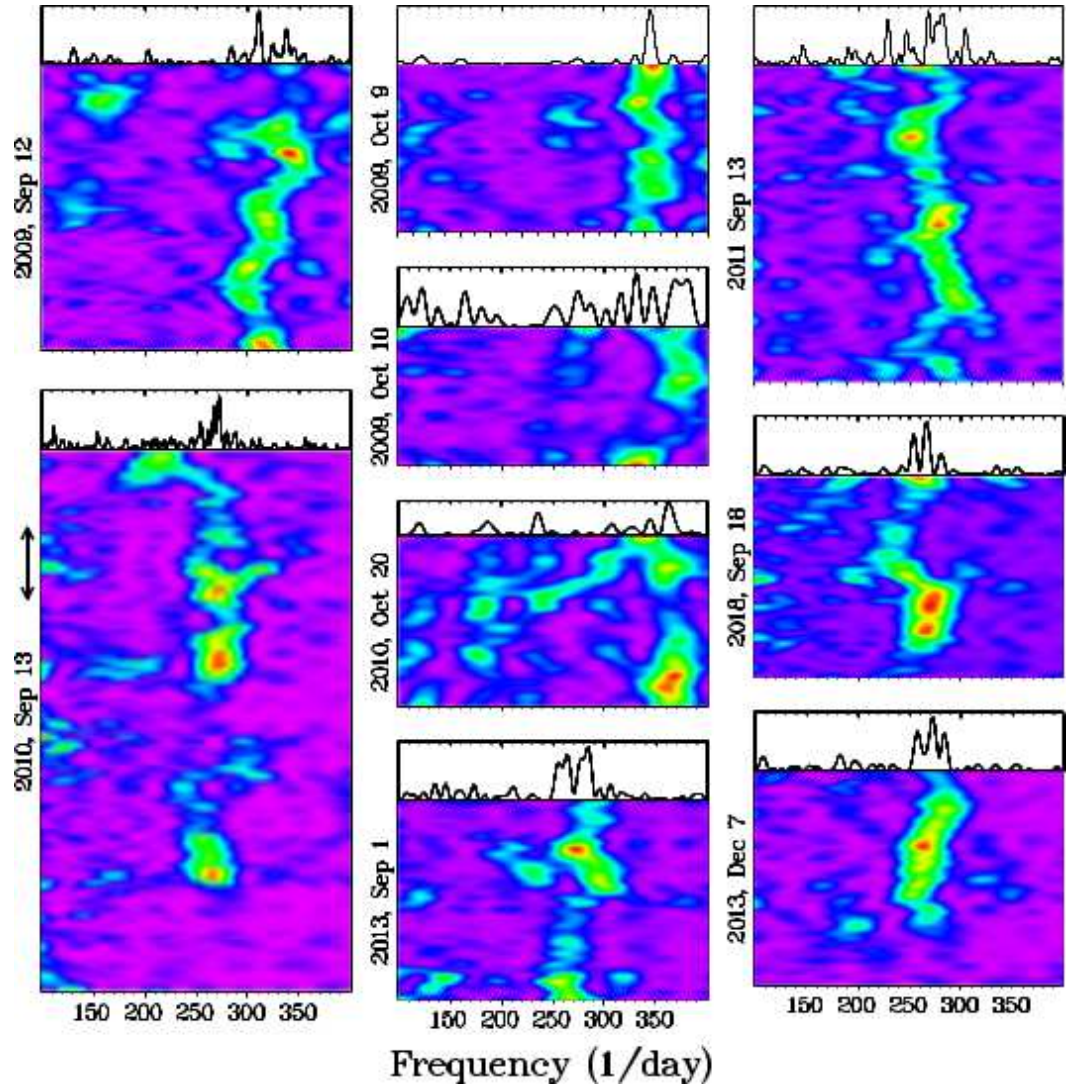}}
      \caption[]{Stacked power spectra of some light curves of
                 V455~And, illustrating the evolution of the non-radial
                 pulsations over time.
                 Time increases from bottom to top. The double arrow at the
                 left, corresponding to 36~min, indicates the length of data
                 sections used to calculate the power spectra. Above each
                 spectrum the conventional Lomb-Scargle periodogram, using
                 the entire light curve, is shown.}
\label{v455and-stacked-1}
\end{figure}

Conventional Fourier techniques, using long light curves, cannot reveal
the duration and evolution of specific pulsations over time.
Stacked power spectra permit a more detailed view of their
evolution over the time-scale of hours. For this purpose I selected some
nightly light curves, the periodograms of which contain comparatively strong
signals in the frequency range of the non-radial pulsations. Periodograms of
sections of these data trains, 36~min long, were constructed, allowing for
an overlap of 90\% between subsequent sections. The individual power spectra
were then stacked on top of each other, resulting in a 2D representation
(frequency vs.\ time; for a more detailed description 
of this technique, see Bruch, 2014). 
The results are shown in Fig.~\ref{v455and-stacked-1}. The
double arrow at the left indicates the length of the data sections used to
calculate the individual spectra. Thus, any vertical structures in the 2D
images smaller that this are not independent from each other.

\begin{figure}
   \parbox[]{0.1cm}{\epsfxsize=14cm\epsfbox{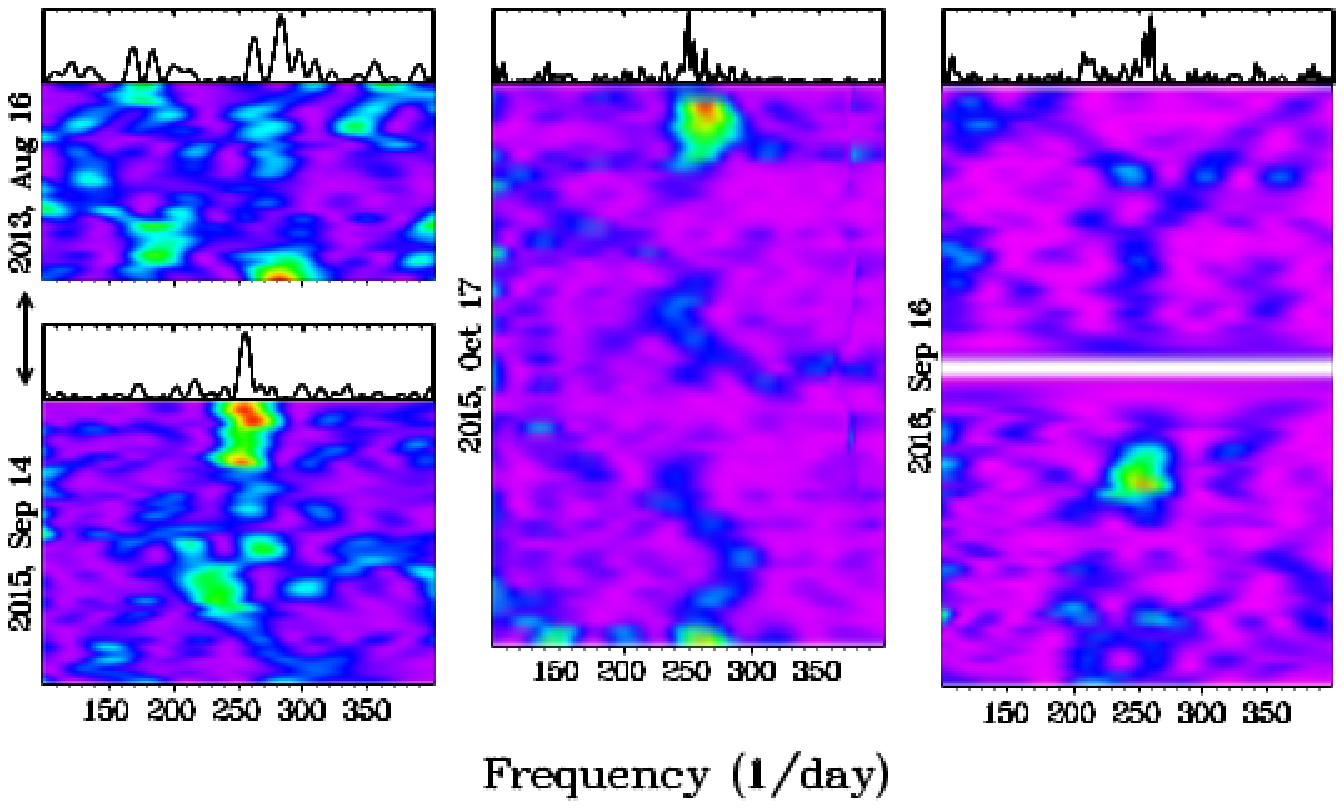}}
      \caption[]{Stacked power spectra of some light curves of
                 V455~And, illustrating cases in which pulsations occuring
                 during short time intervals generate strong signals in
                 conventional periodograms. The figure is organized in the
                 same way as Fig.~\ref{v455and-stacked-1}. (The gap in the
                 spectrum at the right reflects a gap in the light curve.)}
\label{v455and-stacked-2}
\end{figure}

The figure shows that pulsations can persist over several hours, sometimes
with a systematic shift in frequency and, in general, considerable variations
of the amplitude. This can lead to multiple peaks in the conventional
periodograms, complicating their interpretation severely. Moreover, isolated
events in the light curve, lasting for only a short time, can generate
strong signals in the periodogram which may easily be mistaken as evidence
for persistent pulsations. Some examples are shown in
Fig.~\ref{v455and-stacked-2}. It is therefore not astounding that so far
no systematics have been detected in the occurrence of these pulsations
which would permit a better characterization.

\subsection{The spin variations}
\label{V455 And: The spin variations}

In the high frequency regime, V455 And exhibits a stable oscillation with
a period of 67.62~s and its harmonic at half this value, first seen by
Araujo-Betancor et al.\ (2005). It is accompanied by a slightly incoherent
signal at 67.24~s. Their frequency difference corresponds to the 3.5~hr
spectroscopic period. The stable period is identified with the white
dwarf rotation, while G\"ansicke (2007) interprets the 67.24~s period as
due to the illumination of structures in the inner disk.
These variations have also been studied spectroscopically by
Bloemen et al.\ (2013), and photometrically by Silvestri et al.\ (2012) and,
more thoroughly, by Mukadam et al.\ (2016).

The present data cannot add
much to this. Of the 116 light curve only 18 have a time resolution sufficient
to resolve the white dwarf spin period (and none to resolve its harmonic).
Most of them were observed in 2016-18.
A significant signal is seen in the periodogram of all suitable light
curves at an average period of $67.304 \pm 0.020$. This is in between the
spin period and its satellite period. Tests with artificial signals at the
two periods, sampled in the same way as the real light curves, show that the
capacity to resolve them in the periodograms depends on their amplitude ratio
as well as their relative phases. Mukadam et al.\ (2016) observed a ratio of
3-4 for the amplitudes of the 67.24 and 67.62~s signals. In the present data
such a ratio inhibits the resolution of the two periods and shifts the
corresponding periodogram peak to a value slightly longer than 67.24~s.
Thus, the present results are entirely consistent with results obtained in
earlier observing seasons.

\section{SS~Cygni}
\label{SS Cyg}

As the prototype dwarf nova and one of the brightest members of its class
SS~Cyg is arguably one of the best studied of all cataclysmic variables.
Even so, a detailed look at available observations may shed some additional
light on the structure and behaviour of the system. Here, I will
investigate the shape of orbital variations in the light curve and its
supposed variations, and the reality of alleged $\approx$12~min oscillations
associated with the white dwarf rotation.

SS~Cyg has a comparatively long orbital period. The most accurate available
value of 0.27512973~d is based on spectroscopic
measurements of Hessman et al.\ (1984). Although published three and a
half decades ago, the formal error amounts to a phase uncertainty of only
about 0.02 to the present day, which is negligible in the context of this
study.

The data used here consist of 89 $V$ band light curves, all observed during
quiescence, retrieved from the AAVSO International Database. They refer to
the time interval between 2005 and 2015, and have a duration between 1 and 10~h
at a time resolution of 22 to 183~s.

\subsection{Orbital variations}
\label{SS Cyg: Orbital variations}

Orbital variations in SS~Cyg are often not as obvious
as in other systems and may be hardly significant in an individual light curve
in the presence of strong flickering. However, when folding various light
curves on the orbital period, a waveform with two humps of different height
appears. This has first been noted by Voloshina \& Lyutyi (1983) and has also
been documented by Bartolini et al.\ (1985), Voloshina (1986), 
Bruch (1990), Voloshina \& Lyutyi (1993), Kjurkchieva et al.\ (1998) 
and Voloshina \& Khruzina (2000). The large
amount of the data investigated here permit a more rigourous characterization
of these variations.

\begin{figure}
   \parbox[]{0.1cm}{\epsfxsize=14cm\epsfbox{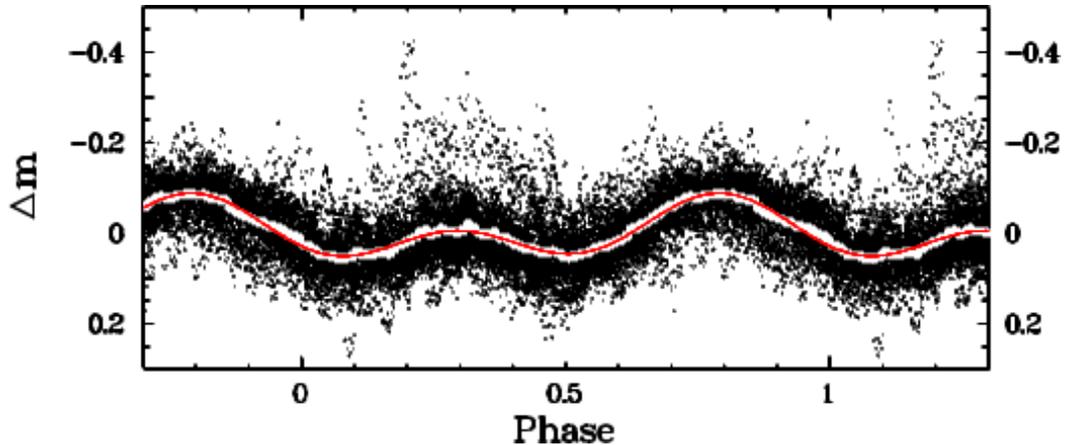}}
      \caption[]{All light curves of SS~Cyg folded on the orbital period.
                 The white ribbon represents a binned version of the same
                 data, using phase bins of width 0.01. The red curve is a
                 two component least squares sine fit with periods fixed to
                 1 and 0.5. Phase 0 corresponds to the inferior conjunction
                 of the red dwarf.}
\label{sscyg-fold-all}
\end{figure}

For this purpose, all light curves were folded on the orbital ephemeris
of Hessman et al.\ (1984).
Combining them into a single phase folded light curves provided
a first approximation of the orbital waveform. This is slightly blurred because
subtracting simply the nightly average magnitude from the original light
curve (Sect.~\ref{The data}) does 
not only remove night-to-night variations, but also a part of the orbital 
variations if the length of a light curve is not an integer multiple of the 
orbital period. However, it is good enough to provide a correction which
depends on the phase coverage of each light curve and which was then applied.
The combined phase folded light curves after this correction provided a cleaner
waveform which is shown in Fig.~\ref{sscyg-fold-all} (black dots). The white
ribbon represents a binned version of the same data, using bins of width 0.01
in phase. Phases are defined such that the zero point corresponds to the
inferior conjunction of the red dwarf.

The variations can well be described by the superposition of two
sine waves with periods equal to the orbital period and half that value (red
graph in the figure), leading to a waveform with two humps of unequal
height separated by minima of almost equal depth. The slightly deeper
minimum occur at phase 0.08, just after the conjunction of the stellar
components. The separation of the minima is 0.425 in phase which means
that the second (higher) hump is somewhat broader than the first. Its
total amplitude is 0.1~mag. All these properties agree well with the
mean light curve shown in fig.~2 of Voloshina \& Khruzina (2000).

The main hump has it maximum at phase 0.79, close to where in many high
inclination CVs a maximum is seen, interpreted as being caused by the changing
visibility of the point of impact on the accretion disk of the stream of matter
transferred from the secondary (i.e., the hot spot). Thus, it is close at
hand to interpret the entire light curve as a superposition of variable
visibility of the hot spot and some other structure which (Bitner et al., 2007)
identify with ellipsoidal variations of the secondary star (with
expected maxima of equal height at phases 0.25 and 0.75). However, this
cannot be the whole story because in that case due to the decreasing
contribution of the cool secondary star at shorter wavelengths the
phase 0.25 maximum should be considerably fainter in the $B$ compared to
the $V$ band, and
even more so in the $U$ band, contrary to the almost constant amplitude in
all bands observed by Voloshina (1986), Voloshina \& Lyutyi (1993) and
Voloshina \& Khruzina (2000). Moreover, double humped orbital light curves
are also observed in short period CVs where the secondary does not contribute
perceptibly to the visual light and ellipsoidal variations can therefore not
explain the waveform. Examples are V455~And (see Fig.~\ref{v455and-orbit-sh})
and WZ~Sge (e.g., Patterson et al., 2018).

Claims have been put forward by Voloshina (1986) and 
Voloshina \& Lyutyi (1993)
concerning a dependence of the shape of the orbital variations on
the outburst phase of SS~Cyg as well as the presence of narrow eclipses
just before and after outbursts. These are based on a comparatively small
number of light curves, meaning that a non-random distribution of strong
flickering flares may mimic variability.

\begin{table}
	\centering
	\caption{Amplitudes of a two component sine fit to the orbital
                 variations of SS~Cyg with periods fixed
                 to the orbital period $P_{\rm orb}$ and to $P_{\rm orb}/2$
                 measured in different years.}
\label{Table: SS Cyg: Amplitudes}

\begin{tabular}{lcccc}
\hline

Year & A ($P_{\rm orb}$)  &  A ($P_{\rm orb}/2$) & $N_{\rm LC}$$^\ast$ &
                                               $\Delta T$$^{\ast \ast}$  \\
     & (mag)            & (mag)              &   & (h)  \\
\hline
 2005 & 0.012 & 0.049 & \phantom{0}4 & \phantom{00}9.5 \\
 2006 & 0.056 & 0.056 &           34 &           138.7 \\
 2007 & 0.055 & 0.028 & \phantom{0}5 & \phantom{0}27.1 \\
 2009 & 0.005 & 0.060 &           16 & \phantom{0}78.0 \\
 2010 & 0.034 & 0.034 & \phantom{0}8 & \phantom{0}21.7 \\
 2011 & 0.035 & 0.031 & \phantom{0}6 & \phantom{0}14.4 \\
 2014 & 0.058 & 0.042 & \phantom{0}4 & \phantom{0}10.0 \\
 2015 & 0.071 & 0.038 &           11 & \phantom{0}61.4 \\
\hline
$N_{\rm LC}$$^\ast$      & \multicolumn{4}{l}{number of light curves} \\
$\Delta T$$^{\ast \ast}$ & \multicolumn{4}{l}{total time base of light curves} \\
\end{tabular}
\end{table}

\begin{figure}
   \parbox[]{0.1cm}{\epsfxsize=14cm\epsfbox{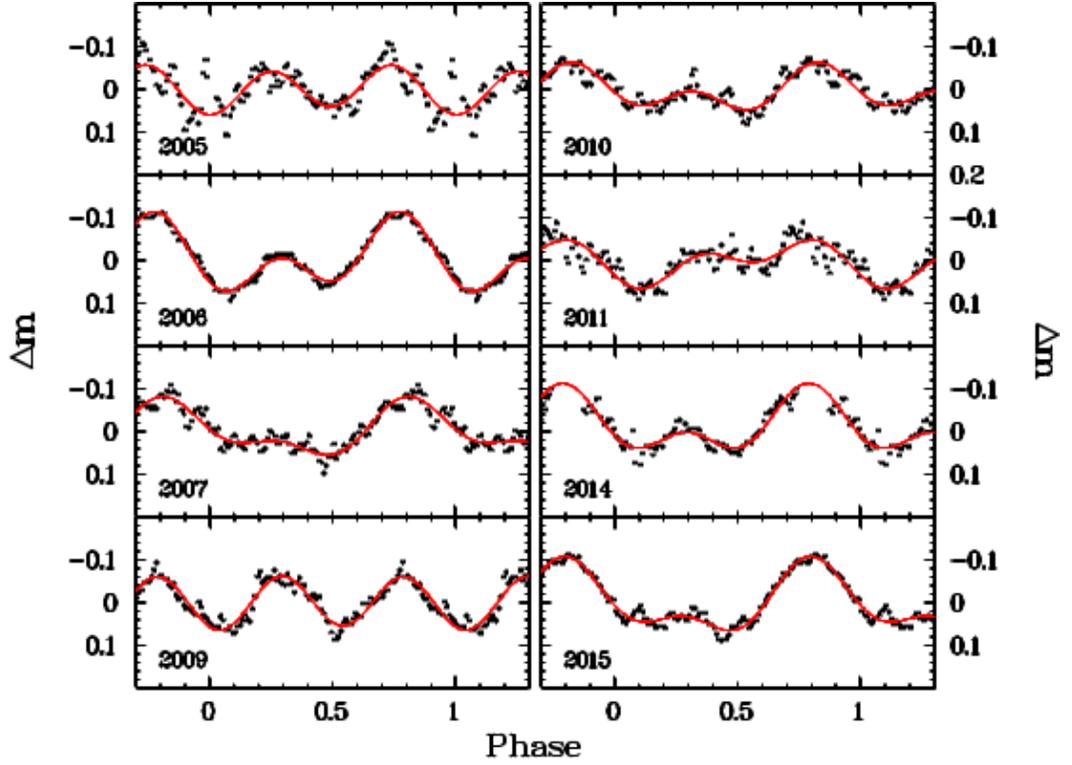}}
      \caption[]{Light curves of SS~Cyg folded on the orbital period and
                 binned, separately for different observing seasons. The
                 red curves are least squares fits with periods fixed to
                 the orbital period and half that values.}
\label{sscyg-fold-years}
\end{figure}

To address this question, I first investigate the average waveform as a
function of observing season. This is done in Fig.~\ref{sscyg-fold-years}
where the binned average orbital light curves in different years are shown
together with the respective two component sine fit (red), and in
Table~\ref{Table: SS Cyg: Amplitudes} where the amplitude of the sine
components are quoted together with the number of contributing light
curves and the total time base of the data.
The latter may provide a feeling of the statistical
reliability of the results. Concentrating on those years with many data
(and thus a well defined average waveform), i.e.,
2006, 2009, and 2015, it is seen that the waveform indeed varies over
time: In 2006 the main hump is about twice as high as the secondary hump;
in 2009 both humps are of approximately equal height; and in 2015 the
fainter hump is all but absent. Similar variations, albeit less clear,
can also be seen in other years.

\begin{figure}
   \parbox[]{0.1cm}{\epsfxsize=14cm\epsfbox{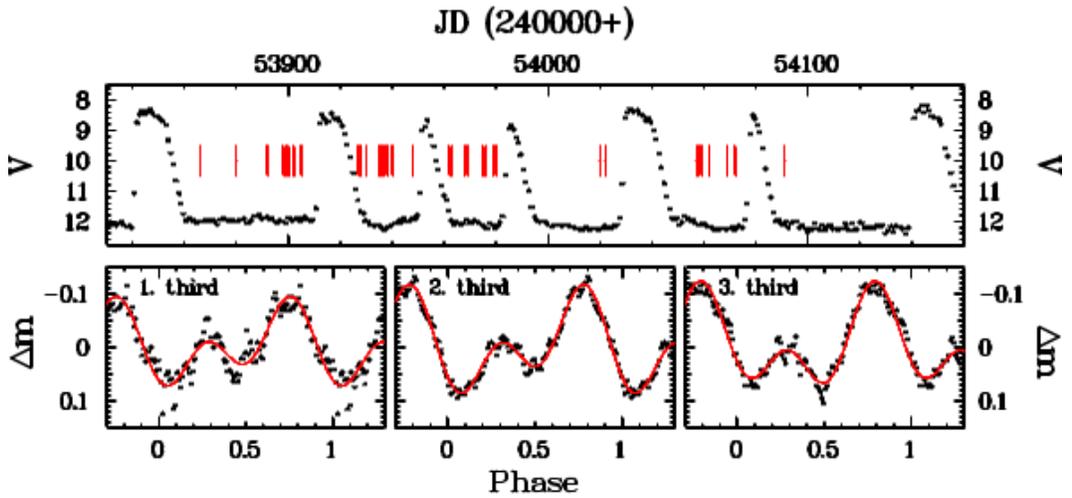}}
      \caption[]{{\it Top:} AAVSO light curves of SS~Cyg during the 2006
                 observing season, binned into 1~d intervals. The red
                 vertical lines indicate the epochs of the nightly light
                 curves used here.
                 {\it Bottom:} Average waveform of the orbital variations
                 in the first, second and last third of the inter-outburst
                 intervals.}
\label{sscyg-fold-thirds}
\end{figure}

These results show that one must not mix data of various observing seasons
in order to investigate dependencies of the waveform (and the presence of
eclipses) on the phase in the outburst cycle. Therefore, I restrict myself
for this purpose to the year with the largest number of available light
curves, i.e., 2006. The AAVSO light curve of that observing season, binned
in 1~d intervals, is reproduced in Fig.~\ref{sscyg-fold-thirds} (top), where
the red vertical bars indicate the epochs of the time resolved light curves
used here. Orbital phase folded light curves falling into the first, second
and last third of each inter-outburst interval were averaged and are shown in
the lower frame of the figure. No systematic differences are seen. There are
indeed some low points in the average waveform of light curves observed just
after outbursts at a phase consistent with the supposed eclipses seen by
Voloshina \& Lyutyi (1993). However, these can all be trace to one particular 
light curve (containing several gaps, arousing the suspicion of mediocre 
observing
conditions), while several others covering the same phases do not contain
this feature. Thus, the present data cannot confirm the presence of eclipses
or variations of the orbital wave form as a function of phase in the outburst 
cycle.

\subsection{SS Cyg at higher frequencies}
\label{SS Cyg: SS Cyg at higher frequencies}

Although not generally recognized as a magnetic system, claims have
repeatedly been put forward in the past for SS~Cyg to be an intermediate
polar. An extensive list of arguments supporting this idea is enumerated by
Giovannelli \& Sabau-Graziati (2012) (see also references cited in that
paper). One of these is the presence of variations with a period of
12.18~min which Giovannelli \& Sabau-Graziati (2012) suspect to be the beat
period between the white dwarf spin and the orbit. It has been observed
by Bartolini et al.\ (1985) and is also mentioned in a theseis by
V.~Tramontana, cited by Giovannelli \& Sabau-Graziati (2012), but unfortunately
not readily available for verification. On the other hand, Bruch (1990) and
Voloshina \& Lyutyi (1993) could not confirm the reality of this signal.

Verifying this issue, using the more extensive data set studied here,
the respective frequency range in power spectra of all light were
investigated. In no case evidence for a periodicity close the 12~min was seen.

\begin{figure}
   \parbox[]{0.1cm}{\epsfxsize=14cm\epsfbox{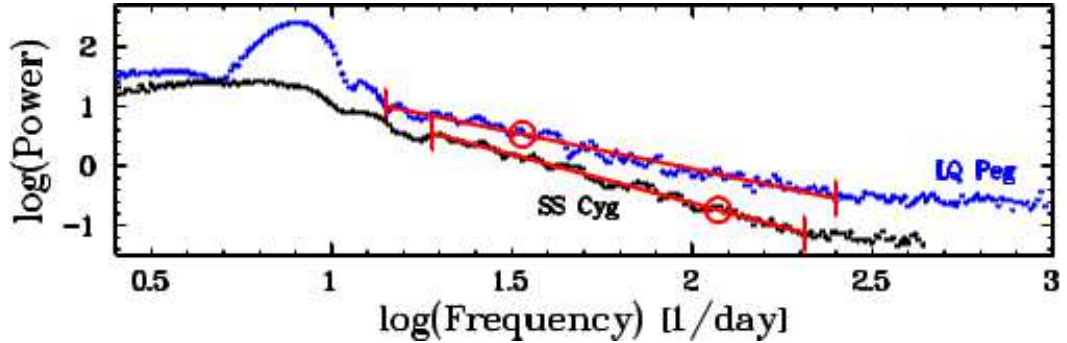}}
      \caption[]{Average power spectra of SS~Cyg and LQ Peg (the latter
                 shifted upwards for clarity) on the double logarithmic
                 scale. The red vertical bars indicate the limits of the
                 red noise realm used to measure the spectral index
                 $\gamma$, i.e., the slope of the red graph. The open circles
                 mark the frequency of the supposed 12.18~min oscillations
                 in SS~Cyg and the centre of the range of QPOs suspected in
                 LQ~Peg.}
\label{lqpeg-powerspectrum}
\end{figure}

At high frequencies the power spectra of cataclysmic variable light curves are
characterized by red noise caused by flickering, meaning that
$P \propto f^{-\gamma}$, where $P$ is the power, $f$ the frequency and $\gamma$
the spectral index. At very high frequencies white noise due to
(approximately) Gaussian measurement errors takes over. On the double
logarithmic scale, red noise causes a linear drop of the power with
increasing frequency with a slope of $-\gamma$, and white noise results in a
constant power level. Distinctive peaks or humps superposed
upon this simple shape then indicate the presence of coherent or quasi
periodic oscillations (QPOs) in a light curve.

Fig.~\ref{lqpeg-powerspectrum} shows the average power spectrum of all
light curves, binned into intervals of with 0.01 in $\log(f)$.
The frequency of the supposed 12.18~min oscillations is
marked by a red open circle. The figure confirms the absence of any
clear signal at this frequency. Instead, the shape of the powerspectrum
confirms the expectations: A linear decline is seen over a wide range.
Within the bounds indicated by the red vertical marks (thus, avoiding the
transition to the white noise realm at very high frequencies) the spectral
index is measured to be $\gamma = 1.62 \pm 0.18$, where the error is
the standard deviation of $\gamma$ derived separately for the various
observing seasons covered by the data.

\section{AQ~Mensae}
\label{AQ Men}

The most basic parameter of any binary system is its orbital period. In
most cases it is more or less straight forward to be measured. The most
reliable methods are the determination of the radial velocity curve and
-- if the orbital inclination is suitable -- the measurement of epochs of
well defined eclipses. Often the light curves also exhibit more
gradual brightness variations which reveal the orbital period. However, in
binary sytems with a complicated structure such as CVs permanent or transient
features are sometimes observed which mimic orbital variations and can thus
lead to false claims of orbital periods. One such case may be AQ~Men.

AQ~Men is a fairly bright ($V \approx 14.3 - 15.3$) CV
which so far has attracted little attention. The star was detected in the
Edinburgh-Cape Blue Object Survey as EC~05114-7955 by Chen et al.\ (2001).
They suspected a dwarf nova nature but did not exclude the possibilty of
AQ~Men being a novalike variable. Armstrong et al.\ (2013)
prefer a classification as a novalike variable, based on the absence of
observed outbursts even many years after discovery, and on the compatible
spectrum.

From radial velocity variations Chen et al.\ (2001) derived a period of
$P_{\rm sp} = 0.130 \pm 0.014$~d. Considering that the total time base
of their observations covers only 1.5 periods the quoted error may be
underestimated. Moreover, the authors did not show a radial velocity curve,
making it difficult to assess the reliability of the derived period.
Within the error, the spectroscopic period is consistent with periods found
in extensive
photometric observations by Armstrong et al.\ (2013). They suspected the
presence of grazing eclipses in the light curve of AQ~Men, recurring with a
period of $P_{\rm ecl} = 0.141471$~d. These eclipses are so shallow that in most
cycles they cannot readily be distinguished in the presence of considerable
flickering activity and only appear more clearly in a phase folded light
curve. Armstrong et al.\ (2013) also found a photometric variation with
a period of $P_{\rm nSH} = 0.13646$~d and a full amplitude of $\sim$0.07~mag 
(see fig.~3 of Armstrong et al., (2013). The period being slightly shorter 
than the eclipse (and thus orbital) period, the authors consider this 
modulation as being due to a negative superhump. Finally, 
Armstrong et al.\ (2013) claim the presence of a longer period of 3.78~d 
with a full amplitude of $\sim$0.12~mag (their fig.~3).

\begin{figure}
   \parbox[]{0.1cm}{\epsfxsize=14cm\epsfbox{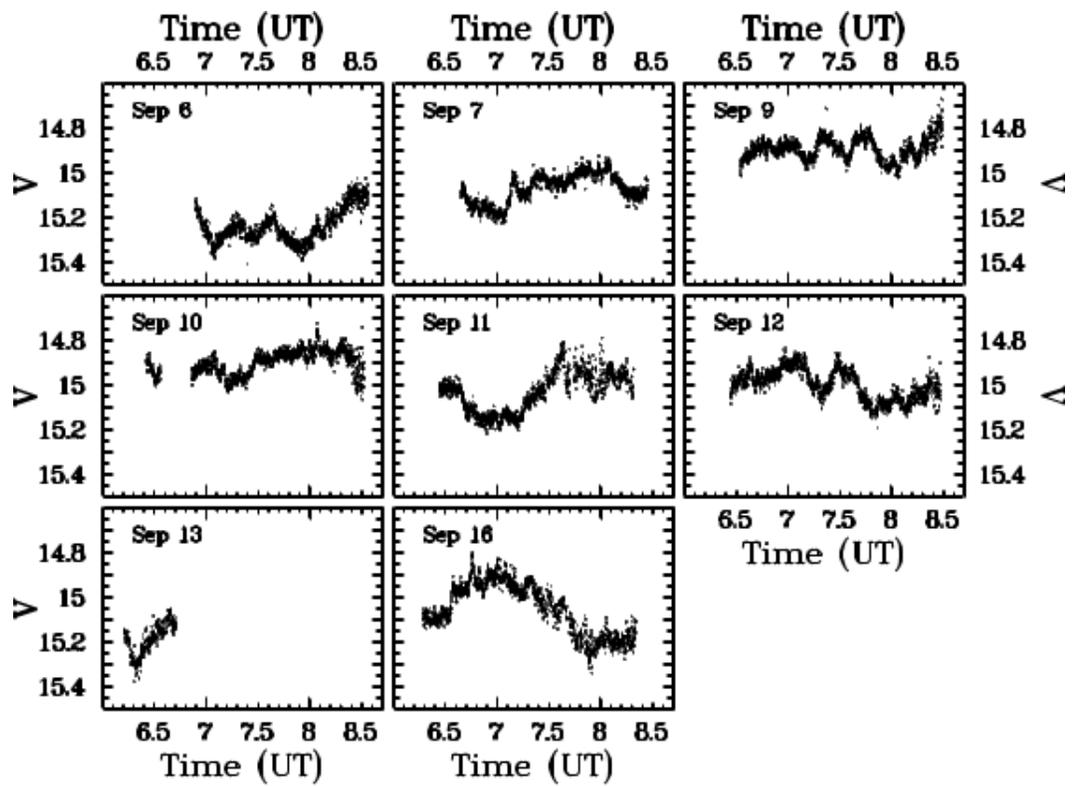}}
      \caption[]{Light curves of AQ~Men observed in 8 nights in 2018,
                 September, all drawn on the same time and magnitude scale.}
\label{aqmen-lightcurves}
\end{figure}

I used the 60~cm Boller \& Chivens telescope of the Observat\'orio do
Pico dos Dias, Brazil, to observe AQ~Men in 8 nights between 2018,
September 6 and 16. Light curves in unfiltered light spanning between 100
and 126~min (except for one shorter run of 30~min) were obtained at a time
resolution of 5~sec. Synthetic aperture photometry was performed on the
original images taken with a blue sensitive CCD IKon-L936-BEX2-DD. 
Magnitudes were measured relative to the primary comparison
star UCAC4 051-003195 (Zacharias et al., 2013) ($V = 14.931$).
For cataclysmic variables the throughput of the instrumentation corresponds
roughly to $V$ (Bruch, 2018). The light curves are shown in
Fig.~\ref{aqmen-lightcurves} where
the time and magnitude scales are the same for all frames. They are
characterized by flickering, variations on the time-scale of the length of
each observing run, and night-to-night variations.

\subsection{Night-to-night variations}
\label{AQ Men: Night-to-night variations}

\begin{figure}
   \parbox[]{0.1cm}{\epsfxsize=14cm\epsfbox{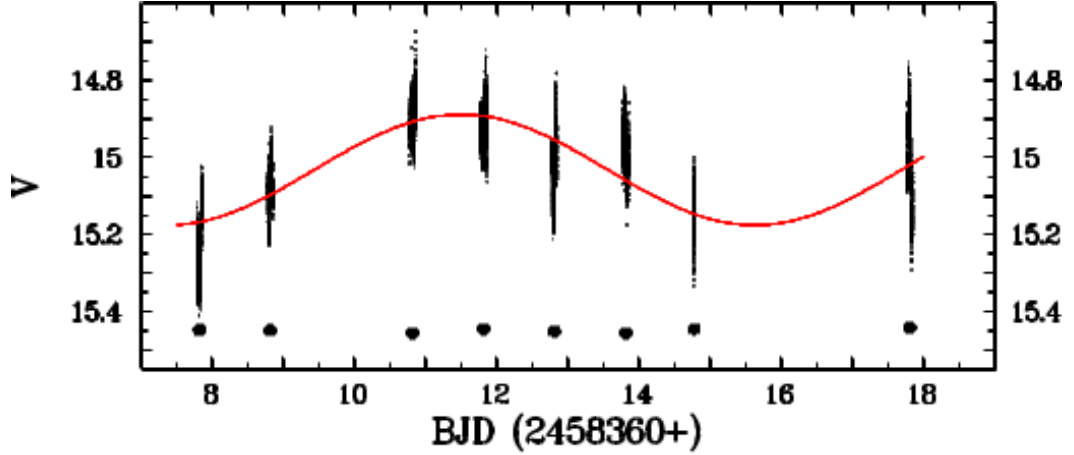}}
      \caption[]{Combined light curves of AQ~Men of 2018,
                 September. The red curve is a least squares sine fit to the
                 data. The filled circles represent the nightly averages of
                 the magnitude difference between the comparison star to
                 AQ~Men and a check star (offset in magnitude by an
                 arbitrary constant).}
\label{aqmen-longterm-lc}
\end{figure}

The night-to-night variations are not random but appear to be systematic.
This is shown in Fig.~\ref{aqmen-longterm-lc},
where the combined light curves of all observing
nights are plotted. The dots beneath the light curve
represent the nightly averages of the difference between the primary
comparison star to AQ~Men and a check star (shifted in magnitude by an
arbitrary constant). It is virtually constant, attesting to the significance
of the night-to-night variations of AQ~Men. A formal least squares sine fit
(red curve in the figure) yields a full amplitude of 0.28 mag and a period
of 8.25~d. However, the data covering a time interval only slightly longer
than this period, it is impossible to say whether this modulation is really
periodic or not. It occurs, in any case, on a time-scale significantly longer
than the 3.78~d variation seen by Armstrong et al.\ (2013) which therefore
cannot be a stable period of AQ~Men. However, a visual inspection of the
upper frame of fig.~1 of Armstrong et al.\ (2013) suggests, at least in the
time interval BJD~2452000 -- 40 a modulation on a similar time-scale as seen
in the present data.

\begin{figure}
   \parbox[]{0.1cm}{\epsfxsize=14cm\epsfbox{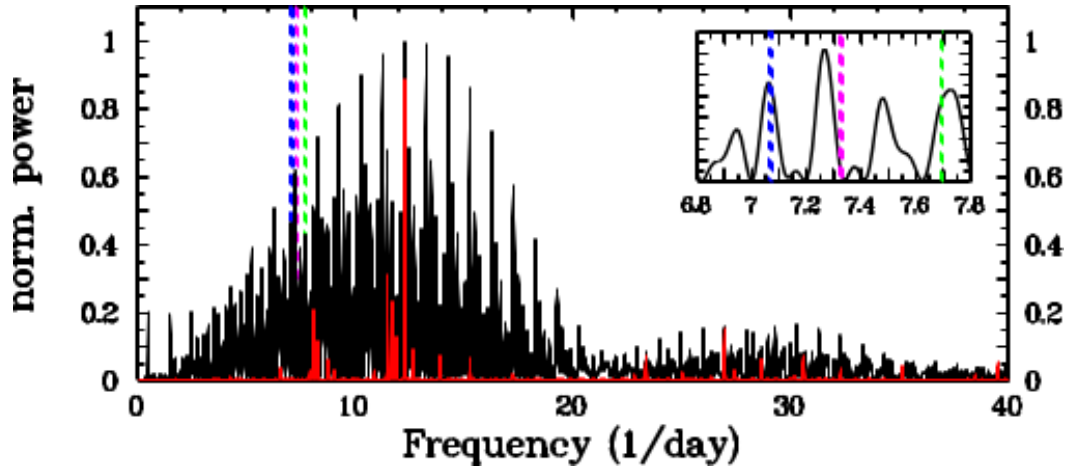}}
      \caption[]{Normalized Lomb-Scargle periodogram (black) and CLEAN power
                 spectrum (red) of the combined light curve of AQ~Men of 2018,
                 September, after subtracting the nightly average.
                 The colour vertical lines indicate the frequency
                 corresponding to the spectroscopic (green), eclipse (blue)
                 and superhump (purple) periods, as quoted in the literature.
                 The insert show a detail of the Lomb-Scargle periodogram
                 around these frequencies.}
\label{aqmen-powerspectrum}
\end{figure}

\subsection{Variations on hourly time-scales}
\label{AQ Men: Variations on hourly time-scales}

\begin{figure}
   \parbox[]{0.1cm}{\epsfxsize=14cm\epsfbox{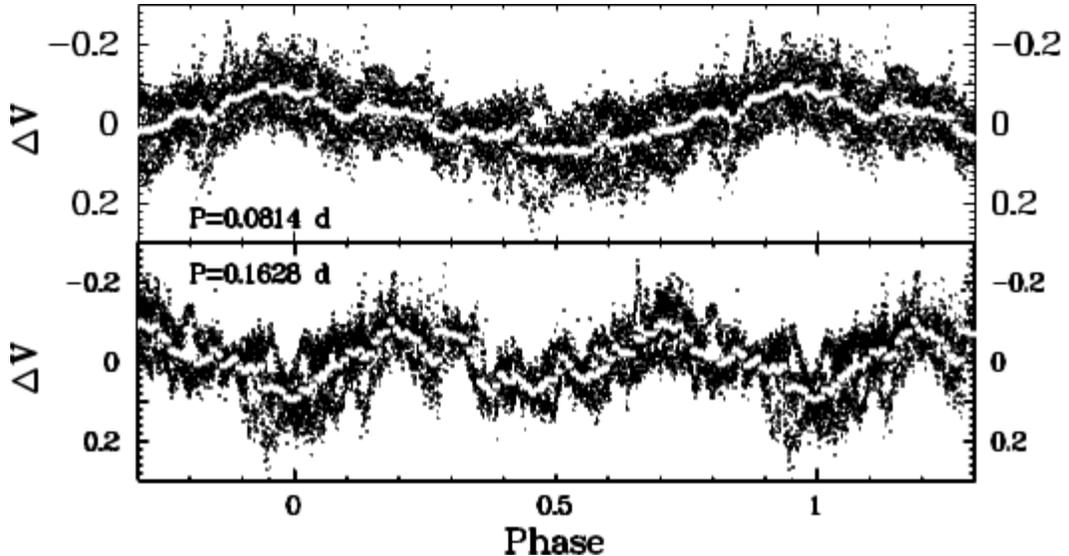}}
      \caption[]{{\it Top:} Combined light curves of AQ~Men of 2018,
                 September, after subtracting the nightly average,
                 folded on the period $P_o = 0.0814$~d. The white ribbon
                 represent the same data, binned in phase intervals of
                 width 0.01.
                 {\it Bottom:} The same, folded on $2 \times P_o$.}
\label{aqmen-folded-lc}
\end{figure}

In order to verify if any of the spectroscopic or photometric periods
seen by Chen et al.\ (2001) and Armstrong et al.\ (2013) can be detected in
the present data, the night-to-night variations were first removed from
the combined data set by subtracting the nightly average magnitudes. A power
spectrum was then calculated. It is shown in Fig.~\ref{aqmen-powerspectrum}
(black graph) and is dominated by the 1~d$^{-1}$ alias patterns of several
frequencies. In order to disentangle the effects of the window spectrum,
the data were also
submitted to the CLEAN algorithm (Roberts et al., 1987). The resulting
spectrum is shown in red in the figure. The highest peak in both, the
Lomb-Scargle periodogram and the CLEAN spectrum occurs at a frequency
of $f_0 = 12.276 \pm 0.031$~d$^{-1}$, where the error (here and in
similar cases in Sections~\ref{LQ Peg: The period: Orbital or superhump} and
\ref{UX UMa: 2015}) is conservatively
defined as the standard deviation of a Gaussian fit to the respective power
spectrum peak in the Lomb-Scargle periodogram.
There are several smaller peaks in the CLEAN
spectrum. However, only the strongest survives if the CLEAN algorithm
is run with different parameters (gain, number of iterations). Therefore, I do
not consider them to be significant. But note that it is by no means
guaranteed that the CLEAN algorithm chooses the correct period among the
aliases (VanderPlas 2018). Thus, the power spectrum suggests that
the light curve of AQ~Men is modulated with a period of $P = 1/(f_0 + n)$ where
$n$ is an integer. Choosing $n=0$ yields $P_0 = 0.0814 \pm 0.0002$~d or
$1.954 \pm 0.005$~h. The light curve, folded on this period,
is shown in Fig.~\ref{aqmen-folded-lc}, where the
white ribbon represent the same data, binned in phase intervals of width 0.01.
A formal sine-fit yields a full amplitude of 0.13~mag for the variations.

How does the period found here relate to the other periods reported to
be present in AQ~Men? Not at all! The frequency corresponding to
$P_{\rm sp}$, $P_{\rm ecl}$ and $P_{\rm nSH}$ are indicated in 
Fig.~\ref{aqmen-powerspectrum}
by green, blue and purple, respectively, broken vertical lines.
As can better be seen in the inset, which contains a blown up version of
the power spectrum in a small range around these frequencies, there is a
satisfactory coincidence of $1/P_{\rm ecl}$ with a minor power spectrum peak,
a marginal coincidence of $1/P_{\rm sp}$ and no coincidence of $1/P_{\rm nSH}$.
Folding the light curves on any of these periods does not lead to convincing
results. I could also not identify any simple arithmetic relationship between
any of these frequencies and $f_0$ or its aliases.

\subsection{Discussion of the findings on AQ Men}
\label{AQ Men: Discussion of the findings on AQ Men}

The distance to AQ~Men derived from the GAIA parallax is $552 \pm 8$~pc
(Bailer-Jones et al., 2018). The interstellar extinction is unknown.
It should be between 0 and the total galactic extinction in the direction
of the star. The reddening maps of Schlafly \& Finkbeiner (2011)
yield a total colour excess in the area around AQ~Men of $E_{B-V} = 0.148$.
Using the standard relation of $A_V/E_{B-V} = 3.1$ this corresponds to a
visual extinction of $A_V = 0.46$. The average magnitude of AQ~Men during
the present observations was $V = 15.0$ (noting that this is only approximate
because the observations were not calibrated). Thus, the absolute visual
magnitude of AQ~Men lies between 5.8~mag (if it suffers the full galactic
extinction) and 6.3~mag (in the absence of any extinction). This is much
brighter than the absolute magnitude of dwarf novae below the 2--3 h gap in
the distribution of orbital periods of CVs (Warner, 1995), meaning that $P_0$
is unlikely to be the revolution period of AQ~Men. However it is compatible
with the faint end of the absolute magnitude distribution of novalike
variables. But then, the bulk of periods of these stars lies above the
period gap. Could it be that the orbital light curve of AQ~Men is double
humped and the true period is $2 \times P_0$, raising it to about 4~h
(considering possible alias values) and thus comfortably into the range
populated by many novalike variables? The lower frame of
Fig.~\ref{aqmen-folded-lc}, showing the
lightcurve folded on $2 P_0$, lends some credibility to this idea because
the minima appear to have different depth. Such double-humped light curves
are not unprecedented. They do not only occur in short period CVs as
mentioned in Sect.~\ref{SS Cyg: Orbital variations} but also in systems
with similar periods such as, for instance
BD~Pav (Barwig \& Schoembs, 1983) ($P_{\rm orb} = 4.30$~h) and
V367~Peg (Woudt et al., 2005) ($P_{\rm orb} = 3.88$~h).
The difference between $2 P_o$ and $P_{\rm sp}$ is 2.3
times the error of the latter which, as mentioned earlier, may have
been underestimated. Both periods may thus be considered compatible.
However, $2 P_o$ is incompatible with $P_{\rm ecl}$. Folding the present data
on $P_{\rm ecl}$ in fact reveals a minimum which looks like an eclipse.
But the evidence is insufficient because only one light curve covers
the respective phase.

Thus, the present results cannot clarify the nature of the various photometric
periods observed in AQ~Men but rather complicate the picture. 

\section{LQ~Pegasi}
\label{LQ Peg}

LQ~Peg (= PG~2133+115) was discovered in the Palomar-Green survey
(Green et al., 1986) and classified as a novalike variable by
Ferguson et al.\ (1984). Most of the time it hovers at a magnitude of
about 14.8~mag, but occasionally low
states occur, as first seen by Sokolov et al.\ (1996) and then by
Watanabe (1999), Kato \& Uemura (1999), Schmidtke et al.\ (2002), and
Kafka \& Honeycutt (2005), testifing to a VY~Scl nature of LQ~Peg.

Not many details about the structure of the system are known. In particular,
no time resolved spectroscopy has been performed. Based on regular brightness
variations Papadaki et al.\ (2006) derived an orbital period of
0.124747~(6). However, even this is contested by Rude \& Ringwald (2012)
who argue that this period is rather due to a negative superhump. They
consider the orbital period to be 0.1342~d. However, it was not
observed directly in their photometry but inferred by two other periods of
0.1425~d (interpreted as a positive superhump) and 
2.37~d (thought to be the precession period of the accretion disk).

In an attempt to shed light upon the question of the orbital period, I
re-analyse here the data of Papadaki et al.\ (2006), downloaded from the 
Centre de Donn\'ee Astronomique de Strassbourg (CDS), consisting of 10 $R$ band
and unfiltered light curves observed in 2004, June and August. These are
complemented by 20 light curves obtained in 2015 August and September, and 17
light curves observed in 2016, August - October, retrieved from the AAVSO
International Database. The latter refer to the $V$ band. The median time
resolution of all light curves is 60~s (range: 44~s -- 125~s).

\begin{figure}
   \parbox[]{0.1cm}{\epsfxsize=14cm\epsfbox{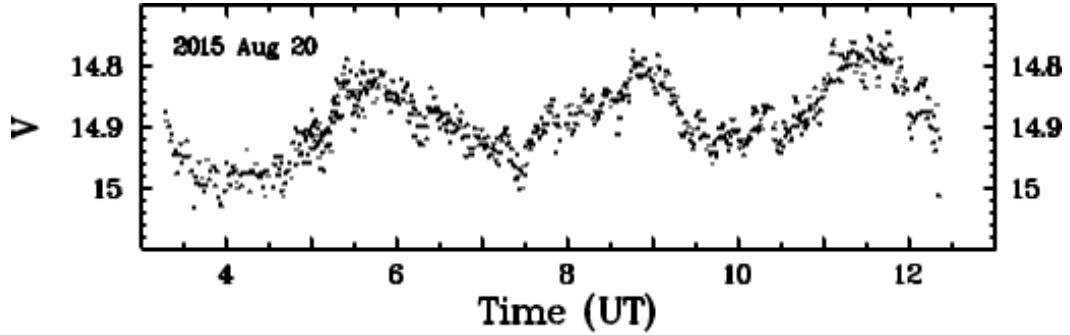}}
      \caption[]{Light curve of LQ~Peg of 2015 August 20, as an example of
                 periodic light modulations in the system.}
\label{lqpeg-lightcurve}
\end{figure}

\subsection{The period: Orbital or superhump?}
\label{LQ Peg: The period: Orbital or superhump}

All light curves of sufficient length clearly exhibit apparently periodic
modulations. Fig.~\ref{lqpeg-lightcurve} shows an example. In order to
investigate their stability and their nature as orbital or superhump
variations the available data sets were divided into three groups according
to the observing seasons mentioned earlier. Since part of the 
Papadaki et al.\ (2006)
light curves were observed in $R$ and others in white light,
the average nightly magnitude was subtracted from the data just as has been
done for the AAVSO data. The combined light curves of the
three observing seasons were subjected to a periodogram analysis.
The periods corresponding to the dominant peaks
are summarized in Table~\ref{Table: LQ Peg: Period table}. 
The periodogram of the 2004 data set contains two closely spaced
alias peaks of very similar height caused by the window function.
The choice between them is possible because the periodograms of the
June and August subsets are free from this complication and -- although
of lower resolution -- exclude within their formal errors one of the alias
frequencies.

\begin{table}
	\centering
	\caption{Periods measured in the light curves of LQ~Peg.}
\label{Table: LQ Peg: Period table}

\begin{tabular}{lccc}
\hline

Dates & N$_{\rm LC}^\ast$ & $\Delta T^{\ast \ast}$ (h) & Period \\
\hline
2004, Jun 1 \ldots Aug 30  & 10 & \phantom{0}44.9 & $0.12474 \pm 0.00004$ \\
2015, Aug 15 \ldots Sep 14 & 20 &           143.5 & $0.12474 \pm 0.00021$ \\
2016, Aug 16 \ldots Oct 16 & 17 &           103.2 & $0.12491 \pm 0.00008$ \\
\hline
\multicolumn{4}{l}{$^{\phantom{\ast} \ast}$ Number of light curves} \\
\multicolumn{4}{l}{$^{\ast \ast}$         Total time base of light curves} \\
\end{tabular}
\end{table}

\begin{figure}
   \parbox[]{0.1cm}{\epsfxsize=14cm\epsfbox{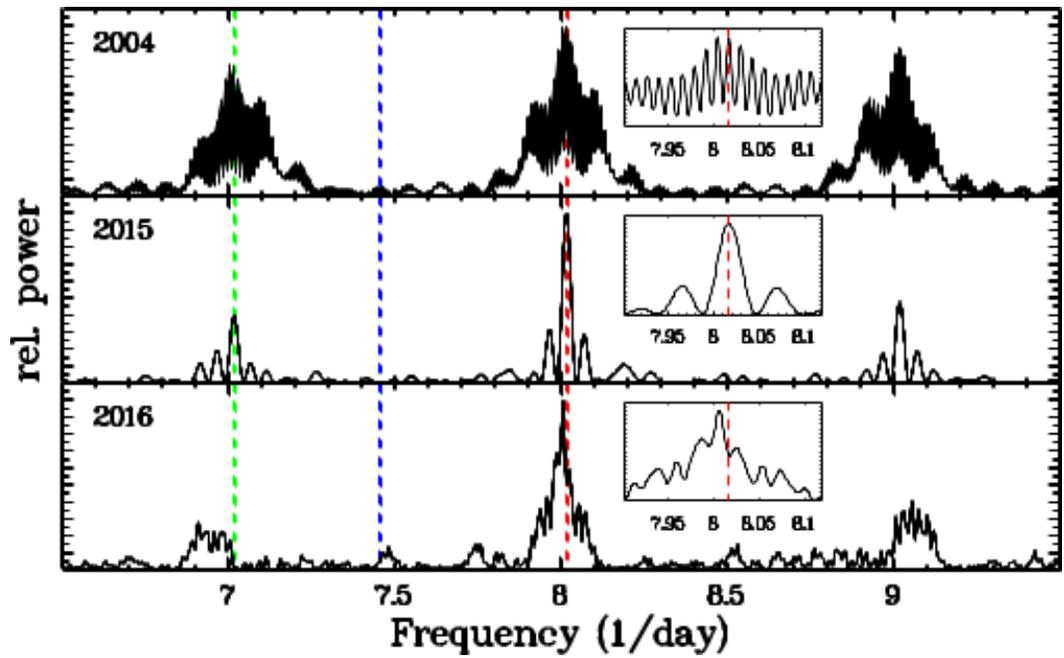}}
      \caption[]{Periodograms of the combined light curves of LQ~Peg
                 obtained in three observing seasons. The inserts
                 contain a blow-up of a small frequency range around
                 dominant peak. The broken vertical lines indicate
                 the frequencies of the period measured by 
                 Papadaki et al.\ (2006) (red, right), and the superhump 
                 (green, left) and orbital (blue, middle) periods suggested 
                 by Rude \& Ringwald (2012).}
\label{lqpeg-periodograms}
\end{figure}

Graphs of the periodograms of the three data groups are shown in
Fig.~\ref{lqpeg-periodograms}. In all cases they are dominated by a 1~d$^{-1}$
alias pattern around a frequency close to the frequency corresponding to
the period $P_{\rm Pap}$ measured by Papadaki et al.\ (2006) which is marked
by the broken red (right) vertical line. The other lines indicate the
frequencies of the superhump $P_{\rm SH}$ (green, left) and orbital period
$P_{\rm orb,RR}$ (blue. middle) proposed by Rude \& Ringwald (2012). The 2004 data are
concentrated in two subgroups separated by more than two months, while
those of the other years are more evenly distributed across the
respective time base. Thus, the alias pattern in the 2004 periodogram
contains more fine structure. This is particularly well visible in the
inserts in Fig.~\ref{lqpeg-periodograms} which contain a blow-up of a
small frequency range around the Papadaki et al.\ (2006) period. Since the
2004 data are the same as investigated by Papadaki et al.\ (2006), it is not
surprising that the frequency $1/P_{\rm Pap}$ is recovered as the highest peak
in the periodogram. In 2015,
the same frequency is dominating the periodogram. However, the dominant
peak in 2016 is slightly offset from $1/P_{\rm Pap}$. While it is still
compatible with $1/P_{\rm Pap}$ on the $2\sigma$ level (see
Table~\ref{Table: LQ Peg: Period table}), tests calculations using different
subensembles of the 2016 data always yielded a peak frequency slightly
smaller than $1/P_{\rm Pap}$, making it unlikely that the difference
is just due to statistical fluctuations. Moreover, folding the combined
data on $P_{\rm Pap}$ reveals clear deviations from a smooth light
curve. This lends some credibility to
the notion that the dominant variations in the light curve of LQ~Pap is
indeed due to a superhump and not to the orbital motion. The superhump
period suggested by Rude \& Ringwald (2012) [the green (left) line in
Fig.~\ref{lqpeg-periodograms}] is then a 1~d$^{-1}$ alias of the true value.
Considering the 2.37 precession period, the revised orbital period would then 
be 0.1185.
However, I do not yet consider this small piece of evidence conclusive.
Moreover, no signal is seen in the periodograms at the orbital frequency
inferred by Rude \& Ringwald (2012) or its revised value.

\begin{figure}
   \parbox[]{0.1cm}{\epsfxsize=14cm\epsfbox{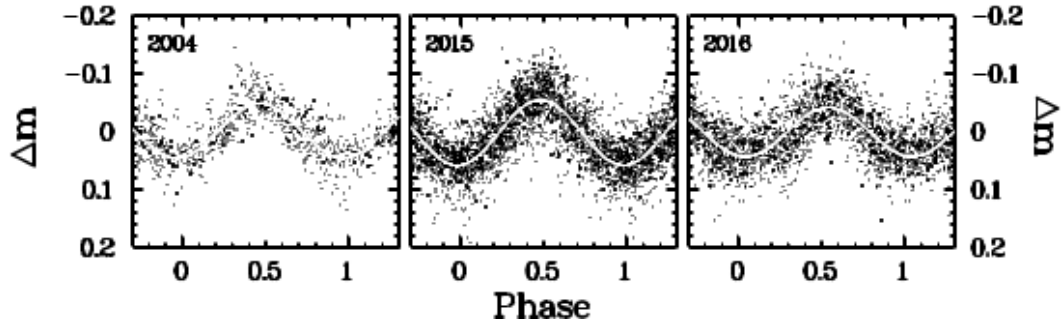}}
      \caption[]{Seasonal light curves of LQ~Peg folded on the dominant
                 period. The white curves are least squares sine fits to
                 the data.}
\label{lqpeg-folded}
\end{figure}

Thus the question whether the strong modulation seen in the light curves
of LQ~Peg is orbital in nature or is a superhump can, unfortunately, not
be settled unequivocally by the present study.
The seasonal data, folded on the period corresponding to the dominant
periodogram peak, are shown in Fig.~\ref{lqpeg-folded}. The full amplitude
of a sine fit to the data (white curves) is 0.09~mag in 2004
and 2016 and at 0.11~mag slightly higher in 2015.

\subsection{The high frequency realm}
\label{LQ Peg: The high frequency realm}

Papadaki et al.\ (2006) suspect a hump to be present in their power spectrum
of LQ~Peg in the frequency range between 26-45~d$^{-1}$, suggesting the
presence of QPOs. They recognize, however, that the signal is too weak to be
certain. The larger number of light curves available here permits to address
this question. The average of all power spectra of LQ~Peg is shown (along
with that of SS~Cyg; see Sect.~\ref{SS Cyg: SS Cyg at higher frequencies})
on a log-log scale in Fig.~\ref{lqpeg-powerspectrum}.
The broad hump at low frequencies
is caused by the orbital (or superhump) variations. The frequency range of
the suspected QPOs extends between $\log(f) = 1.41$ and 1.65. The centre
of this range is marked by an open red circle. There is
no distinct feature in this range. Thus, the existence of the QPOs is not
confirmed.

Instead, the red noise realm, caused by flickering, is well expressed and
extends between $1.15 < \log{f} < 2.4$ (red vertical bars in the figure).
A linear least squares fit (red graph) yields a spectral index of
$1.24 \pm 0.04$ where the error is the standard deviation of $\gamma$
derived from the three observing seasons. There is thus no significant
change of $\gamma$ over time. In contrast, Papadaki et al.\ (2006) quote
very different nominal values. The difference, however, may not be significant
because Papadaki et al.\ (2006) recognize that their $\gamma$ depends on the
exact definition of the fitting interval, leading to a wide uncertainty 
interval.

It should be noted that $\gamma$ is significantly larger in SS~Cyg
compared to LQ~Peg. This indicates a difference in the distribution of
flickering power in the two systems: While in both of them the absolute
value of the power grows at longer time-scales, the relative power of
long lasting flares dominates more strongly over the power on short 
time-scales in SS~Cyg compared to LQ~Peg.

\section{RW Trianguli}
\label{RW Tri}

RW~Tri is a deeply eclipsing novalike variable. The orbital period was
determined by Robinson et al.\ (1991) to be 0.231883297~d. The inclusion of
this system in the present study is motivated by the recent detection by
Smak (2019) of a negative superhump present in light curves of RW~Tri
observed in 1957 and 1994. Moreover, albeit as a novalike variable
RW~Tri is not expected to undergo large magnitude variations (noting
that low states such as those occurring in VY~Scl stars have not been
observed in this system), low amplitude ($\approx$0.5~mag) oscillations
on the time-scale of some tens of days have been reported by
Honeycutt et al.\ (1994) and Honeycutt (2001). The present data permit
also to further explore this aspect.

A dense series of observations have been obtained by AAVSO observers
during 2015-2016. It consists of 62 light curves observed between 2015,
October 18 and 2016, February 9. They refer all to the $V$ band. The total
observing time amounts to 379~h, while the average length of individual light
curves is 4.3~h (range: 1.8~h -- 12.9~h). Their average time resolution is
61~s (range: 39~s -- 86~s). Since I
am interested here in variations other than eclipses, these have been
removed from all data before further processing.

\subsection{Long term variability}
\label{RW Tri: Long term variability}

During a $\approx$250~d interval (out of a total of $\approx$1000~d of
observations) during the 1992-1993 observing season Honeycutt et al.\ (1994)
observed cyclic variations with a total amplitude of 0.45~mag and a period
of 25.1~d (Honeycutt, 2001) in RW~Tri.
This is much more than eventual magnitude zero point problems
in data taken with different instruments (see Sect.~\ref{The data}) can
conceal. Therefore it is worthwhile to investigate the original AAVSO light 
curves (i.e., before subtracting the nightly average magnitude) for similar
variations.

\begin{figure}
   \parbox[]{0.1cm}{\epsfxsize=14cm\epsfbox{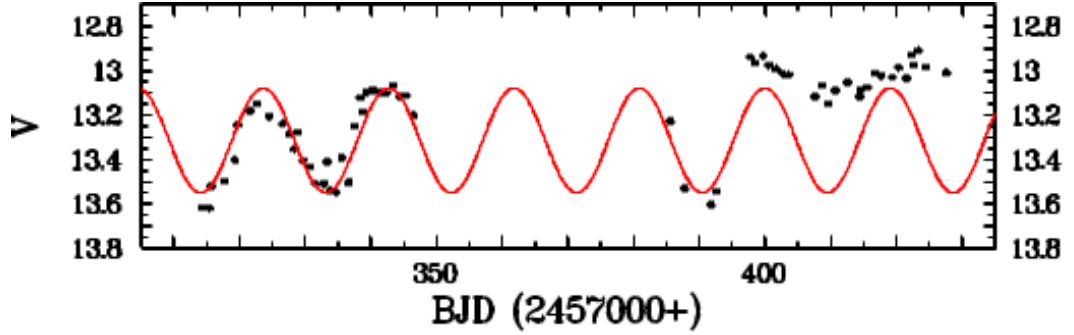}}
      \caption[]{Long term light curve of RW~Tri during the 2015-2016
                 observing season. The black dots represent
                 the average magnitudes of the nightly light curves. The red
                 curve is a least squared sine fit to the data
                 before BJD~2457395.}
\label{rwtri-long-term}
\end{figure}

The larger density of the present data as compared to those of Honeycutt 
et al.\ (1994) permits a more rigorous characterization of the long term 
variations. The black dots in Fig.~\ref{rwtri-long-term} respresent average
magnitudes of the nightly light curve. In particular during the first part of 
the observing season a very clear cyclic variability of the average magnitude
is obvious which is lost when the season draws to an end. The red curve is a
least squares sine fit to the average magnitudes before BJD~2457395. Its total
amplitude of 0.47~mag is very similar to that seen by Honeycutt (2001),
but at $19.10 \pm 0.01$~d the period is somewhat shorter. After BJD~2458395 the
system brightens sightly and the amplitude of the variations decreases
significantly. However, at least for another 19~d cycle their phase is 
maintained.

\subsection{Superhumps in RW~Tri?}
\label{RW Tri: Superhumps in RW Tri?}

Recently, Smak (2019) reported the detection of a negative superhump
at a period of 0.2203~d (i.e., 5\% shorter than the orbital period) in
light curves of RW~Tri observed in 12 nights in September 1994. Some
indications for the superhump were also found in 4 light curves taken in
November-December 1957. If this feature is real and persistent, as it is
for instance in TT~Ari (Bruch 2019), it should easily be revealed in
the present data. However, it is not.

To study this issue, I now employ the RW~Tri light curves after subtraction
of the nightly average magnitude. A Lomb-Scargle periodogram of the combined
data was calculated. Since there is a gap of 40~d in the long term light
curve (see Fig.~\ref{rwtri-long-term}) I also investigated the data before
and after the gap separately. The results were almost identical.

\begin{figure}
   \parbox[]{0.1cm}{\epsfxsize=14cm\epsfbox{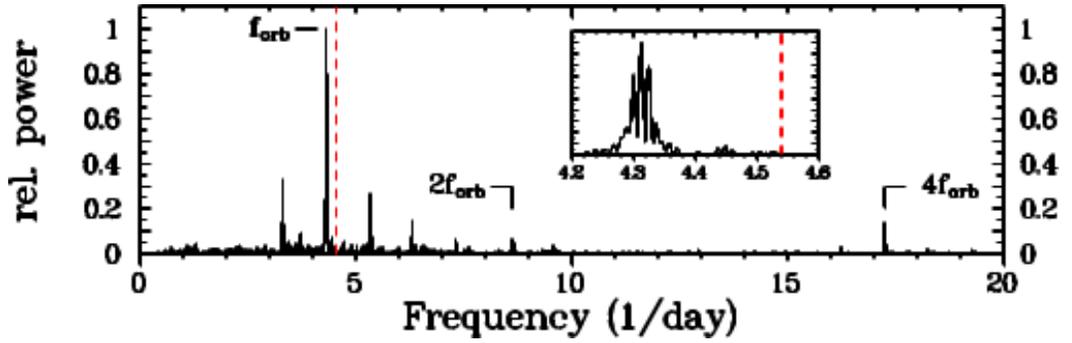}}
      \caption[]{Lomb-Scargle periodogram of the combined light curves
                 of RW~Tri. The broken vertical line indicates the
                 frequency of the negative superhump reported by
                 Smak (2019). The insert shows the region around the
                 superhump and orbital frequencies on an amplified scale.}
\label{rwtri-powerspectrum}
\end{figure}

The power spectrum is displayed in Fig.~\ref{rwtri-powerspectrum}. Significant
peaks appear at the orbital frequency $f_{\rm orb}$, twice this value and,
somewhat surprisingly and with considerable strength, $4f_{\rm orb}$. Several
1~d$^{-1}$ alias of these features are also obvious. All other structures
which may be considered at the brink of significance, can be traced to
only one of the two parts of the total data set and therefore do not reflect
persistent variations in RW~Tri. Indeed, subtracting the average waveform
from the combined data, the power spectrum of the resulting light curve
does not contain convincing evidence for a residual periodic signal.

In particular, no trace of a superhump is present. Its frequency, as measured
by Smak (2019) is marked by a broken vertical line in
Fig.~\ref{rwtri-powerspectrum}. The absence of the superhump signal is even
more evident in the insert which shows the corresponding
spectral region, also including the orbital frequency, on an amplified scale.
The satellite peaks of the orbital maximum can be explained as beat
frequencies with the total observing window.

Thus, it must be concluded that at least during the 2015-2016 observing
season no evidence of a negative superhump in RW~Tri exists.

\section{UX Ursae Majoris}
\label{UX UMa}

UX~UMa is another well known deeply eclipsing novalike variable. At
0.196671278~d (Baptista et al., 1995) it has an orbital period only slightly
smaller than RW~Tri. I include UX~UMa in this study because based on an
extensive observing campaign in 2015 de Miguel et al.\ (2016) recently found
a modulation with a period of 3.680~d in its light curve which they
interpret as being due to a retrograde precession of the accretion disk.
An associated negative superhump at the beat period of the precession and
the orbit is also seen at 0.186700~d. At the end of the discussion of
their observations de Miguel et al.\ (2016) remark (citing): {\it So it's a
decent bet, though by no means sure, that these superhump effects have
been lurking, unsuspected, in many previous observations of UX~UMa.} As
I will show below, they loose their bet. At least, the presence of
superhumps is not a common phenomenon in UX~UMa, but rather an exception.

I use a total of 198 light curves covering a total time base of 1133 h,
observed during several observing seasons. They were all retrieved from
the AAVSO International Database and refer to the $V$ band. The eclipses
were removed. A summary of these observations is given in
Table~\ref{Table: UX UMa: Summary}, where the number of light
curves and the total time base is listed as a function of observing season.
The majority of the observations is concurrent with those analysed by
de Miguel et al.\ (2016) and -- although they did not provide a detailed list
of their observations, so that a direct comparison is not possible -- I
suspect that at least some (or even most) of the data studied here are the
same as those used by them. Since these light curves cannot provide much
information beyond the more detailed investigation of de Miguel et al.\ (2016),
the 2015 data are only briefly treated in Sect.~\ref{UX UMa: 2015} in order
to verify the previous results, before the data of other observing seasons
are regarded in Sect.~\ref{UX UMa: Other observing seasons} in the
quest for the presence of superhumps at other epochs.

\begin{table}
	\centering
	\caption{Summary of UX~UMa light curves per observing season}
\label{Table: UX UMa: Summary}

\begin{tabular}{lcc}
\hline

Year & $N_{\rm LC}$$^\ast$ & $\Delta T$$^{\ast \ast}$ (h)  \\
\hline
 2005 & \phantom{00}7 & \phantom{0}26.8 \\
 2012 & \phantom{00}5 & \phantom{0}31.9 \\
 2013 & \phantom{0}17 & \phantom{0}96.3 \\
 2014 & \phantom{0}23 &           193.7 \\
 2015 &           112 &           649.5 \\
 2016 & \phantom{0}10 & \phantom{0}38.2 \\
 2017 & \phantom{0}13 & \phantom{0}40.8 \\
 2018 & \phantom{0}11 & \phantom{0}52.3 \\
\hline
\multicolumn{3}{l}{$^{\phantom{\ast} \ast}$ Number of light curves} \\
\multicolumn{3}{l}{$^{\ast \ast}$         Total time base of light curves} \\
\end{tabular}
\end{table}

\subsection{2015}
\label{UX UMa: 2015}

\begin{figure}
   \parbox[]{0.1cm}{\epsfxsize=14cm\epsfbox{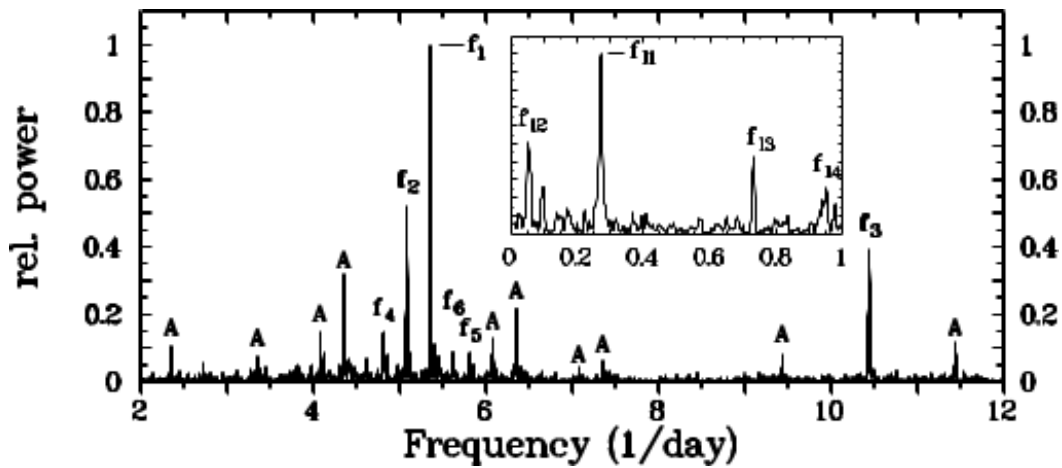}}
      \caption[]{Lomb-Scargle periodogram of the combined light curves
                 of UX~UMa. The main significant peaks are identified and
                 discussed in the text. Obvious 1~d$^{-1}$ aliases are
                 labelled as ``A''. The insert contains the low frequency
                 part of the periodogram of the combined data before
                 subtracting the nightly average magnitude.}
\label{uxuma-2015-ps}
\end{figure}

The analysis of the 2015 data fully confirms the results of 
de Miguel et al.\ (2016). The are summarized in Fig.~\ref{uxuma-2015-ps} and
Table~\ref{Table: UX UMa: 2015 power spectrum}. The figure (main frame) shows
the intermediate frequency range of the power spectrum of the combined light
curves of UX~UMa after subtraction of the nightly average magnitude, while
the insert is a blow-up of the low frequency part, using the data before
subtracting the average. The table lists the frequencies (and periods) of
the main power spectra peaks (in the order of their relative strength).

At low frequencies, two independent signals are seen. $f_{l1}$ is the prominent
variation interpreted by de Miguel et al.\ (2016) as due to disk precession. The
second signal at $f_{l2}$ (corresponding to a period of 18.8~d) appears to
be somewhat clearer in the present data than, e.g., in fig.~2 of
de Miguel et al.\ (2016) who suspected it to be spurious. I am more inclined to
consider the reality of a period or quasi-period close to 19~d an open
issue. $f_{l3}$ and $f_{l4}$ are obviously caused by a beat of $f_{l1}$ and
$f_{l2}$ with the window spectrum which has a strong maximum at 1~d$^{-1}$.

\begin{table}
	\centering
	\caption{Frequencies (in d$^{-1}$) and periods (in d) in the 2015
                 power spectrum of UX~UMa}
\label{Table: UX UMa: 2015 power spectrum}

\begin{tabular}{ll}
\hline

\multicolumn{2}{l}{Low ($<$ 1d$^{-1}$) frequency range:} \\ [1ex]
\phantom{x}$f_{l1} = \phantom{0}0.2719 \pm 0.0039 \phantom{\equiv 1 - f_{l1}}$ &
$P_{l1} = \phantom{1}3.676\phantom{00} \pm 0.053$ \\
\phantom{x}$f_{l2} = \phantom{0}0.0531 \pm 0.0035 \phantom{\equiv 1 - f_{l1}}$ &
$P_{l2} = 18.82\phantom{000} \pm 1.2$ \\
\phantom{x}$f_{l3} = \phantom{0}0.7399 \pm 0.0036 \equiv 1 - f_{l1}$ &  \\
\phantom{x}$f_{l4} = \phantom{0}0.9491 \pm 0.0038 \equiv 1 - f_{l2}$ &  \\
[1ex]
\multicolumn{2}{l}{Intermediate (2--12d$^{-1}$) frequency range:} \\ [1ex]
\phantom{x}$f_{1} = \phantom{0}5.3573 \pm 0.0046 \equiv \phantom{2}f_2 - f_{l1}$ &  \\
\phantom{x}$f_{2} = \phantom{0}5.0835 \pm 0.0036 \phantom{\equiv 2f_2 - f_{l1}}$ &
$P_{2} = \phantom{0}0.196715 \pm 0.00014$ \\
\phantom{x}$f_{3} = 10.4059 \pm 0.0052 \equiv 2f_1 - f_{l1}$ &           \\
\phantom{x}$f_{4} = \phantom{0}4.8123 \pm 0.0059 \equiv \phantom{2}f_2 - f_{l1}$ &  \\
\phantom{x}$f_{5} = \phantom{0}5.8146 \pm 0.0053 \equiv \phantom{2}f_2 + f_{l1}$ &  \\
\phantom{x}$f_{6} = \phantom{0}5.6179 \pm 0.0030 \equiv \phantom{2}f_1 + f_{l1}$ &  \\

\hline

\end{tabular}
\end{table}

At first glance, the power spectrum at intermediate frequencies (main frame
of Fig~\ref{uxuma-2015-ps}) appears confusing. However, it is really quite
simple. Apparently significant signals are labelled $f_1$ -- $f_6$, while
obvious 1~d$^{-1}$ aliases are identified by an ``A''. The main peak at $f_1$ is
the beat between the precession frequency and the orbital frequency $f_2$,
i.e., the negative superhump frequency. All other signals can be interpreted
as sums or differences between the orbital and the precession frequency, as
detailed in Table~\ref{Table: UX UMa: 2015 power spectrum}.
Thus, disregarding $f_{l2}$, the light curve of UX~UMa contains only two
independent frequencies: the orbital and the precession (or, alternatively,
the superhump) frequency.

\subsection{Other observing seasons}
\label{UX UMa: Other observing seasons}

\begin{figure}
   \parbox[]{0.1cm}{\epsfxsize=14cm\epsfbox{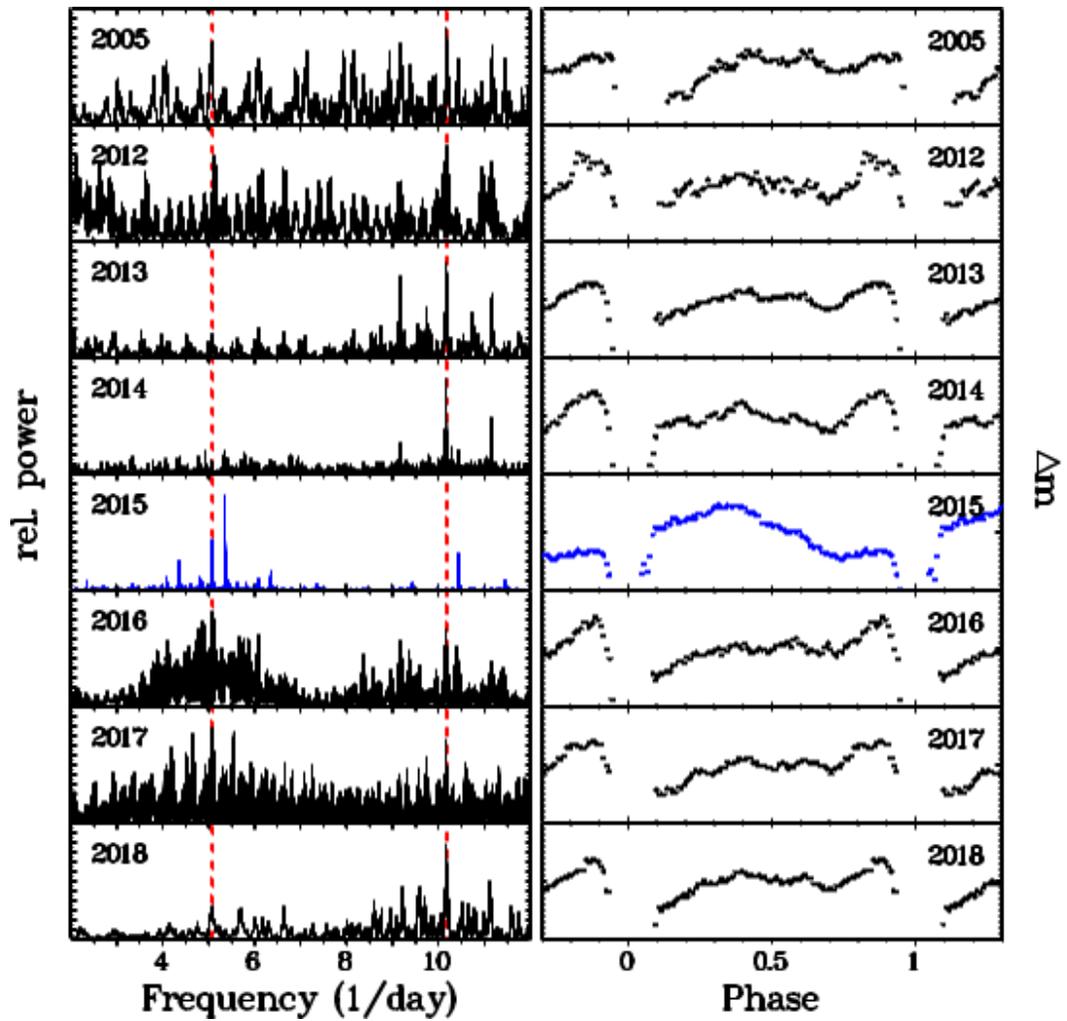}}
      \caption[]{{\it Left:} Lomb-Scargle periodograms of the combined light
                 curves of UX~UMa in different years. The broken vertical
                 lines indicate the orbital frequency and its first overtone.
                 {\it Right:} The combined seasonal light curves folded on
                 the orbital period, binned in intervals of with 0.01 in
                 phase.}
\label{uxuma-multi-ps}
\end{figure}

To which degree the behaviour of UX~UMa observed in 2015 repeats itself
at other epochs? In order to answer this question, Fig.~\ref{uxuma-multi-ps}
(left column) contains the power spectra of the combined light curves observed 
in 8 different seasons [including the 2015 season (in blue) to facilitate
the comparison]. The orbital frequency and its first overtone is indicated
by the red broken vertical lines. It must be taken into account that the number
of light curves in all other years is considerably smaller than in 2015. Thus,
non-periodic variations together with the random distribution of data within
the observing season, cause a multitude of structures in the power spectrum
and any real periodic signal will be less outstanding. Significant peaks
(together with their 1~d$^{-1}$ aliases) are
always (except in 2015) seen at the first harmonic of the orbital frequency
$f_2$ and in general, but often at a reduced strength, at $f_2$ itself. In no
year other than 2015 any indication of the negative superhump at $f_1$ is
seen. In 2016, a peak close to $f_3$ appears. However, the absence of any
convincing signal at $f_1$ -- so much stronger than $f_3$ in 2015 -- and
the presence of many other peaks of similar strength makes it doubtful that
it really reflects the presence of a superhump.

The right column of Fig.~\ref{uxuma-multi-ps} contains the seasonal light
curves folded on the orbital period, binned in phase intervals of width 0.01.
The average orbital waveform consists of a more or less well defined
hump encompassing roughly the first 2 thirds of the orbit, followed by a
stronger hump, canonically considered as the manifestation of the hot spot
in high inclination CVs. Only in 2015 the waveform is drastically different
(see also fig. 3 of de Miguel et al., 2016).

Thus, both, the power spectra and the average orbital waveform, clearly
indicate that the photometric behaviour of UX~UMa observed in 2015 is not
seen during any other observing season investigated here. It must therefore
be regarded as exceptional. This evidently leads to the question of the
reason for the emergence of the unusual light curve in 2015 and, assuming
that its interpretation by de Miguel et al.\ (2016) is correct, the change of
structure of the accretion disk. This question, however, will not be answered
here.

\section{Conclusions}

The purpose of this paper is to verify and expand on previous published 
details of miscellaneous variations observed in some cataclysmic variables.
It also shows the value of archival data, which are readily available but
have not yet been exploited adequately. The main findings can be summarized
as follows:

{\em V455 And:} Improved orbital ephemeries are presented. Claims of a 
positive superhump during quiescence cannot be substantiated since the 
corresponding variations can be explained as a aliasing effect with orbital
variations. On the other hand, the continuous presence of negative superhumps
is confirmed. The variations of the frequencies of non-radial pulsations of
the white dwarf over the years suggest that in 2008 the star was still too 
hot after the 2007 outburst to exhibit pulsations and that it has reached
its equilibrium temperature by 2012. Pulsations are not stable but can
rapidly appear and disappear or change their frequencies.

{\em SS Cyg:} The waveform of the double humped orbital variations changed 
from year to year in the sense that the fainter hump occuring during the
first half of the orbit may all but vanish or may almost reach the
amplitude of the stronger second maximum. Previous claims for a dependence
of the waveform on the phase within the outburst cycle of SS~Cyg cannot be
confirmed. The same is true for 12~min oscillations which have been claimed 
to be evidence for an intermediate polar nature of SS~Cyg.

{\em AQ Men:} Variations of the system brightness with a period of $\approx$8
days (which, however, requires confirmation) are seen. A modulation of the
light curve with a period of about 2~h (or twice this value) is also observed
Its nature is not clear. None of the brightness changes of AQ~Men could be 
related to previously determined photometric or spectroscopic periods.

{\em LQ Peg:} Conflicting interpretations in the literature concerning
cyclic modulations in this star as being orbital or superhump variations
could not unambiguously be resolved. 
QPOs have been suspected previously to occur in LQ~Peg. The present
observations do not contain any indication for their presence.

{\em RW Tri:} A negative superhump claimed to have been observed in 1994 and
possibly in 1957 is definitely not a permanent feature since no trace could
be seen in extensive data obtained during 2015-16. On the other hand, the
occasional occurrence of cyclic variations on the time-scale of 
$\approx$20~d is confirmed at a much higher confidence level than permitted
by previous reports.

{\em UX UMa:} The unusual photometric behaviour of UX~UMa, exhibiting
a negative superhump and variations on the precession period of a warped
accretion disk in 2015, is confirmed. But it is also shown that this
behaviour is unusual and does not repeat in none of the seven other observing
seasons for which data are available.

\section*{Acknowledgements}

This work is partly based on observations taken at the Observat\'orio do
Pico dos Dias operated by the Laborat\'orio Nacional de Astrof\'{\i}sica,
Brazil. It also depends to a great deal on archival data downloaded from
the AAVSO International Database. I thank the numerous dedicated observers 
of the AAVSO for their contributions, together with the AAVSO staff for 
maintaining the database. Without their efforts this work would not have 
been possible. This research made also use of the VizieR catalogue access 
tool, CDS, France (DOI: 10.26093/cds/vizier).

\section*{References}

\begin{description}
\parskip-0.5ex

\item   Araujo-Betancor, S., G\"ansicke, B.T., Hagen, H.-J., et al., 
        2005, A\&A, 430, 629
\item   Armstrong, E., Patterson, J., Michelsen, E., et al., 
        2013, MNRAS, 435, 707
\item   Bailer-Jones, C.~A.~L., Rybizki, J., 
        Fouesneau, M., Mantelet, G., Andrae, R., 2018, AJ, 156, 58
\item   Baptista, R., Horne, K., Hilditch, R.W., Mason, K.O., Drew J.E.,
        1995, ApJ, 448, 395
\item   Bartolini, C., et al., 1985, Mbga.conf,  50,
\item   Barwig H., Schoembs R., 1983, A\&A, 124, 287
\item   Bitner, M.A., Robinson, E.L., Behr, B.B., 2007, ApJ, 662, 564
\item   Bloemen, S., Steeghs, D., De Smedt, K., et al., 2013, MNRAS, 429, 3433
\item   Bruch, A., 1990, AcA, 40, 369
\item   Bruch, A., 2014, A\&A, 566, A101
\item   Bruch, A., 2018, NewA, 58, 53
\item   Bruch, A., 2019, MNRAS, 489, 2961
\item   Chen, A., O'Donoghue, D., Stobie, R.S., Kilkenny, D., Warner, B.,
        2001, MNRAS, 325, 89
\item   de Miguel, E., Patterson, J., Cejudo, D., et al., 2016, MNRAS, 457, 1447
\item   Eastman, J., Siverd, R., Gaudi, B.S., 2010, PASP, 122, 935
\item   Ferguson, D.H., Green, R.F., Liebert, J., 1984, ApJ, 287, 320
\item   G{\"a}nsicke, B.T., 2007, ASPC,  597, ASPC..372
\item   Giovannelli, F., Sabau-Graziati, L., 2012, MmSAI, 83, 698
\item   Green, R.F., Schmidt, M., Liebert, J., 1986, ApJS, 61, 305
\item   Hagen, H.-J., Groote, D., Engels, D., Reimers, D., 1995, A\&AS, 111, 195
\item   Hessman, F.V., Robinson, E.L., Nather, R.E., Zhang, E.-H.,
        1984, ApJ, 286, 747
\item   Honeycutt, R.K., 2001, PASP, 113, 473
\item   Honeycutt, R.K., Robertson, J.W., Turner, G.W., Vesper, D.N.,
        1994, ASPC, 277, 56
\item   Kafka, S., Honeycutt, R.K., 2005, IBVS, 5597, 1
\item   Kato, T., Imada, A., Uemura, M., et al., 2009, PASJ, 61, S395
\item   Kato, T., Uemura, M., 1999, IBVS, 4786, 1
\item   Kjurkchieva, D., Marchev, D., Drozdz, M., 1998, Ap\&SS, 262, 453
\item   Kozhevnikov, V.P., 2015, NewA, 41, 59
\item   Lomb, N.R., 1976, Ap\&SS, 39, 447
\item   Maehara, H., Imada, A., Kubota, K., et al., 2009, ASPC,  57, 404
\item   Matsui, R., Uemura, M., Arai, A., et al., 2009, PASJ, 61, 1081
\item   Mukadam, A.S., Pyrzas, S., Townsley, D.M., et al., 2016, ApJ, 821, 14
\item   Nogami, D., Hiroi, K., Suzuki, Y., et al., 2009, ASPC,  52, 404
\item   Papadaki, C., Boffin, H.M.J, Sterken, C., et al., 2006, A\&A, 456, 599
\item   Patterson, J., Stone, G., Kemp, J., et al., 2018, PASP, 130, 064202
\item   Roberts, D.H., Lehar, J., Dreher, J.W., 1987, AJ, 93, 968
\item   Robinson, E.L., Shetrone, M.D., Africano, J.L., 1991, AJ, 102, 1176
\item   Rude, G.D., Ringwald, F.A., 2012, NewA, 17, 453
\item   Scargle, J.D., 1982, ApJ, 263, 835
\item   Schlafly, E.F., Finkbeiner, D.P., 2011, ApJ, 737, 103
\item   Schmidtke, P.C., Ciudin, G.A., Indlekofer, U.R., et al.,
        2002, ASPC,  539, 26
\item   Silvestri, N.M., Szkody, P., Mukadam, A.S., et al., 2012, AJ, 144, 84
\item   Smak, J., 2019, AcA, 69, 79
\item   Sokolov, D.A., Shugarov, S.Y., Pavlenko, E.P.,
        1996, ASSL,  219, 208
\item   Szkody, P., Mukadam, A.S., G\"ansicke, B.T., et al., 2013, ApJ, 775, 66
\item   VanderPlas, J.T., 2018, ApJS, 236, 16
\item   Voloshina, I.B., 1986, PAZh, 12, 219
\item   Voloshina, I.B., Khruzina, T.S., 2000, ARep, 44, 89
\item   Voloshina, I.B., Lyutyi, V.M., 1983, PAZh, 9, 612
\item   Voloshina, I.B., Lyutyi, V.M., 1993, ARep, 37, 34
\item   Warner, B., 1995, {\it Cataclysmic Variables Stars}, Cambridge 
        University Press
\item   Watanabe, T., 1999, SVOLJ Var.\ Star Bull., 34, 3
\item   Woudt, P.A., Warner, B., Spark, M., 2005, MNRAS, 364, 107
\item   Zacharias, N., Finch, C.T., Girard, T.M., et al., 2013, AJ, 145, 44

\item      Allen, C.W. 1973, Astrophysical Quantities, third edition 
      (Athlone Press: London)
\item      Baptista, R. 2012, Mem.S.A.It., 83, 530
\item      Bateson, F.M. 1974, Publ.\ Var.\ Star Sect., RASNZ, 2, 1
\item      Brown, A.G.A, Vellenari, A., Prusti, T., et al. 2016, A\&A, 595, A2
\item      Bruch, A. 1993, 
      MIRA: A Reference Guide (Astron.\ Inst.\ Univ.\ M\"unster
\item      Bruch, A. 2016, New Astr., 46, 90
\item      Bruch, A. 2017a, New Astr., 52, 112
\item      Bruch, A. 2017b, New Astr., in press
\item      Bruch, A., \& Diaz, M.P. 2017, New Astr., 50, 109
\item      Caceci, M.S., \& Cacheris, W.P. 1994, Byte, May 1984, 340
\item      Catal\'an, S., Isern, J., Garc\'{\i}a-Berro, E., \& Ribas, I. 
      2008, MNRAS 387, 1693
\item      Claret, A., \& Bloemen, S. 2011, A\&A, 529, A75
\item      Dura\v{s}evi\'c, G., Latkovi\'c, O., Vince, I., \& Cs\'eki, A. 2010,
      MNRAS, 409, 329
\item      Dura\v{s}evi\'c, G., Vince, I., Antokhin, I.I., et al. 2012, MNRAS,
      420, 3081
\item      Eastman, J., Siverd, R., \& Gaudi, B.S. 2010, PASP, 122, 935
\item      Eggleton, P.P. 1983, ApJ, 268, 368
\item      Evans, P.A., Beardmore, A.P., Osborne, J.P., \& Wynn, G.A.
      MNRAS, 399, 1167
\item      FitzGerald, M.P. 1970, A\&A, 4, 234
\item      Garrido, H.E., Mennickent, R.E., Dura\v{s}evi\'c, G., et al. 2013,
      MNRAS, 428, 1594
\item      Gicger, A. 1987, Acta Astron., 37, 29
\item      Goliasch, J., \& Nelson, L. 2015, ApJ, 809, 80
\item      Harmanec, P., \& Scholz, G. 1993, A\&A, 279, 131
\item Hoffman, D.J, Harrison, T.E., Coughlin, J.L., et al. 2008, AJ, 136, 1067
\item      H{\o}g, E., Fabricius, C., Makarov, V.V., et al. 2000, A\&A, 355, L27
\item      Houk, N. 1978, Catalogue of two dimensional spectral types for the
      HD stars, Vol.\ 2, University of Michigan
\item Jacoby, G.H., Hunter, D.A., \& Christian, C.A. 1984, ApJ Suppl., 56, 257
\item      Kalomeni, B., Nelson, L., Rappaport, S., et al. 2016, ApJ, 833, 83
\item      Kazarovetz, E.V., Samus, N.N., Durlevich, O.V., Kireeva, N.N.,
      \& Pastukhova, E.N. 2008, IBVS 5863
\item      Kilkenny, D., \& Laing, J.D. 1990, SAAO Circ., 14, 11
\item      Kordopatis, G., Gilmore, G., \& Steinmetz, M. 2013, AJ, 146, 134
\item      Lasota, J.-P. 2001, New Astron.\ Rev., 45, 449  
\item      Mennickent, R.E., \& Djura\v{s}evi\'c, G. 2013, MNRAS, 432, 799
\item      Mennickent, R.E., Dura\v{s}evi\'c, G., Ko{\l}aczkowski, Z., \&
      Michalska, G., 2012, MNRAS, 421, 862
\item      Menzies, J.W., O'Donoghue, D., \& Warner, B. 1986, ApSS, 122, 73
\item      Nauenberg, M. 1972, ApJ, 175, 417
\item      Osaki, Y. 1996, PASP, 108, 39
\item      Pezzuto, S., Bianchini, A., \& Stagni, R. 1996, A\&A, 312, 865
\item Paczy\'nski, B., \& Schwarzenberg-Czerny, A. 1980, Asta Astron, 30, 127
\item      Pojmanski, G. 2002, Acta Astron, 52, 397 
\item      Rafert, J.B., \& Twigg, L.W. 1980, MNRAS, 193, 79
\item      Ritter, H. 2010, Mem.\ S.A.It., 81, 849
\item      Ritter, H., \& Kolb, U. 2003, A\&A, 404, 301
\item      Salaris, M., Serenelli, A., Weiss, A., \& Miller Bertolami, M.
      2009, ApJ, 692, 1013
\item      Shears, J., \& Poyner, G. 2009, JBAA, 120, 169
\item      Smak, J. 1983, ApJ, 272, 234
\item      Szkody, P., \& Mattei, J.A. 1984, PASP, 96, 988
\item      Walter, F., Bond, H.E., \& Pasten, A. 2006, IAU Cir.\ 8663
\item      Warner, B. 1987, MNRAS, 227, 23
\item Warner, B. 1995, Cataclysmic Variable Stars, Cambridge University Press,
      Cambridge
\item      Wilson, R.E. 1979, ApJ, 234, 1054
\item      Wilson, R.E., \& Devinney, E.J. 1971, ApJ, 166, 605
\item      Zacharias, N., Finch, C.T., Girard, T.M., et al.\ 2013, AJ, 145, 44
\item      Zhao, J.K., Oswalt, T.D., Willson, L.A., Wang, Q., \& Zhao, G. 2012,
      ApJ, 746, 144
\item Zorotovic, M., Schreiber, M.R., \& G\"ansicke, B.T. 2011, A\&A, 536, A42
\item      Zwitter, T., \& Munari, U. 1995, A\&AS, 114, 575
\item      Zwitter, T., \& Munari, U. 1996, A\&AS, 117, 449

\end{description}

\end{document}